\documentclass[aps,pre,reprint,superscriptaddress,amsmath,showpacs,amssymb]{revtex4-2}
\usepackage{amssymb}
\usepackage{tipa}
\usepackage{bm,epsfig,mathrsfs,amsmath,amssymb,graphicx}
\usepackage{epsfig,amsmath,graphicx,amssymb}
\usepackage{graphicx}
\usepackage{dcolumn}
\usepackage{bm}
\usepackage{color}
\usepackage{tikz}
\usepackage{tikz-3dplot}
\usepackage{tkz-euclide}

\usepackage[hyperindex,breaklinks,unicode,colorlinks=true,linkcolor=blue,citecolor=blue,urlcolor=blue]{hyperref}
\usepackage[all]{hypcap}
\usetikzlibrary{decorations.markings}
\usepackage[toc,title,titletoc,header]{appendix}

\newcommand{\tps}{\tilde{t}_{p}^*}

\begin{document}
\newcommand\uleip{\affiliation{Institut f\"ur Theoretische Physik, Universit\"at Leipzig,  Postfach 100 920, D-04009 Leipzig, Germany}}
\newcommand\chau{\affiliation{ 
 Charles University,  
 Faculty of Mathematics and Physics, 
 Department of Macromolecular Physics, 
 V Hole{\v s}ovi{\v c}k{\' a}ch 2, 
 CZ-180~00~Praha, Czech Republic
}}

\title{Maximum efficiency of low-dissipation heat pumps at given heating load}
\author{Zhuolin Ye}\email{zhuolinye@foxmail.com}\uleip
\author{Viktor Holubec}\email{viktor.holubec@mff.cuni.cz}\chau

\begin{abstract}
We derive an analytical expression for maximum efficiency at fixed power of heat pumps operating along a finite-time reverse Carnot cycle under the low-dissipation assumption. The result is cumbersome, but it implies simple formulas for tight upper and lower bounds on the maximum efficiency and various analytically tractable approximations. In general, our results qualitatively agree with those obtained earlier for endoreversible heat pumps. In fact, we identify a special parameter regime when the performance of the low-dissipation and endoreversible devices is the same. At maximum power, heat pumps operate as work to heat converters with efficiency 1. Expressions for maximum efficiency at given power can be helpful in the identification of more practical operation regimes.
\end{abstract}

\maketitle

\section{Introduction}

Besides the uneasy transfer to carbon-free electricity generation, e.g., by using solar, wind, water, geothermal, fission, and, soon hopefully also fusion power, a possibility to fight global warming is to use more efficient devices. To this end, practical heat engines can already operate at high efficiencies differing from the reversible efficiency by less than a factor of 2~\cite{Holubec2015}. On the other hand, most state of the art heat pumps can easily decrease energy consumption for heating by a factor of 3 \cite{ammar2019performance}, 
which is still far below their second law theoretical maximum (Carnot) coefficient of performance (COP)
\begin{equation}
\epsilon_{C}=T_{ h}/(T_{h}-T_{c}).
\label{eq:Carnot}
\end{equation}
For example, a common situation in households with room (target) temperature $T_h \approx 293$ K and heat source temperature $T_c \approx 273$ K corresponds to $\epsilon_C \approx 14.7$, i.e. one joule of electric energy can transfer 14.7 joules of heat. The recent raised interest in heat pumps~\cite{chua2010advances, mohanraj2021performance, hu2021effects} is thus fully deserved as already their implementations with current COPs might help to reduce $\rm CO_2$ emissions
\cite{aye2010evaluation, hong2016greenhouse}.

It is well known that the maximum COP~\eqref{eq:Carnot} is attained in heat pumps that operate quasi-statically and, similarly as for heat engines~\cite{PhysRevE.96.062107}, their output power (called heating load) is negligibly small. Heat pumps able to heat a household thus have to operate outside the quasi-static limit, in a regime described by finite-time thermodynamics. For heat engines and refrigerators, similar considerations lead to a thorough investigation of their efficiency at maximum power using a variety of models~\cite{gordon1993performance, rubin1982optimal, chen1989unified,apertet2013efficiency,PhysRevE.88.062115,PhysRevLett.105.150603, gonzalez2020energetic, PhysRevE.101.052124,hernandez2015time,benenti2011thermodynamic, izumida2014work,VandenBroeck2005,Izu2015, Izu-EPL-heat, Izu-EPL-refri,uzdin2014universal, abah2012single, rossnagel2014nanoscale,schmiedl2007efficiency, segal2008stochastic, jarzynski1999feynman,Dechant2017,Holubec2014,PhysRevLett.102.130602,VandenBroeck2005,PhysRevE.89.012129,PhysRevE.91.022136,ye2017universality,PhysRevE.87.012133,PhysRevE.81.041106,PhysRevE.77.041118,PhysRevLett.111.050601,Izu-EPL-heat,Long2018,Izu2015}. However, idealized models of heat pumps, e.g., based on the endoreversible thermodynamics~\cite{leff1978eer,blanchard1980coefficient}, imply diverging maximum power with COP 1. At maximum power, such heat pumps thus operate as pure work to heat converters, which is a highly undesirable operation regime.

As a result, efficiency at maximum power is for heat pumps not a useful measure of performance. A more informative figure of merit is the maximum efficiency at a given power, which generalizes various trade-off measures between power and efficiency~\cite{apertet2013efficiency, yan1990class,de2012optimal,hernandez2001unified,angulo1991ecological,Izu-EPL-refri, de2013low, PhysRevE.82.051101,PhysRevE.93.032152,long2015ecological}. Maximum efficiency at given power was thoroughly studied for various models of heat engines~\cite{PhysRevE.93.050101,PhysRevE.96.012151,PhysRevE.102.012151,See-AppendixC,Whitney2014,Whitney2015,Long2016,holubec2016maximum,Dechant2017} and refrigerators \cite{PhysRevE.101.052124,Long2018,ye2020maximum}. However, besides numerical studies~\cite{guo2020equivalent}, the only available analytical results for heat pumps were obtained for endoreversible heat pumps \cite{blanchard1980coefficient,cheng1995performance,chen1995study}. 

In this paper, we derive the analytical expression for maximum COP at a given heating load for Carnot-type
low-dissipation (LD) heat pumps. In Secs.~\ref{sec-model} and~\ref{sec-define-variables}, we introduce the considered model and define the thermodynamic quantities of interest. In Sec.~\ref{sec-four}, we discuss the performance of the LD heat pumps operating at maximum power. In Sec.~\ref{sec-max-COP-at}, we present our main results. Specifically, the lower and upper bounds on maximum COP at a given power for LD heat pumps are derived in Sec.~\ref{sec-bounds}. And in Sec.~\ref{sec-arbitrary}, we derive a general expression for the maximum COP together with an analytically tractable approximation.  In Sec. \ref{comparison-endo}, we compare the obtained results for maximum COP of LD heat pumps to the known results for endoreversible heat pumps. We conclude in Sec. \ref{sec-conclusion-outlook}.

\label{sec-bounds}
\label{sec-arbitrary}

\section{Model}
\label{sec-model}

\begin{figure}[t]
\centering
\includegraphics[trim={0cm 0cm 0cm 0cm}, width=0.65\columnwidth]{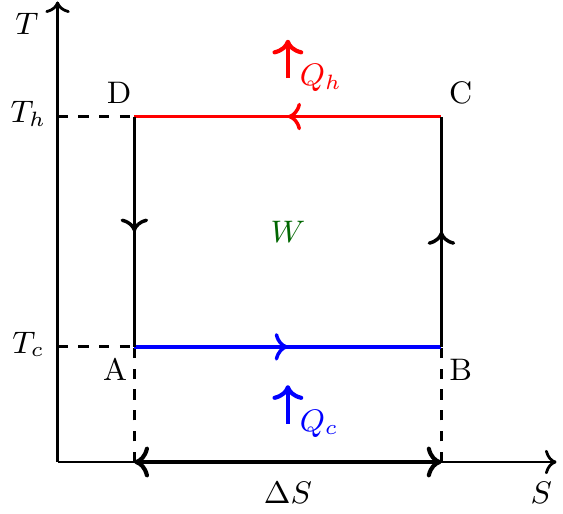}
\caption{Bath temperature $T$-system entropy $S$ diagram of the considered Carnot heat pump cycle. The red (blue) horizontal line denotes the hot (cold) isotherm. The black vertical lines depict the adiabats. Per cycle, the input work $W$ is consumed to pump the heat $Q_c$ from the cold bath at temperature $T_{ c}$ and deliver the heat $Q_h = Q_c + W$ into the hot bath at temperature $T_{ h}$.}
\label{fig:T-S}
\end{figure}

Consider a heat pump operating along the finite-time reverse Carnot cycle depicted in Fig. \ref{fig:T-S}. The cycle consists of two isotherms and two adiabats. During the cold isotherm, the system extracts heat $Q_c$ from the cold bath at temperature $T_{ c}$. Afterward, during the hot isotherm, it uses the input work $W$ to pump this heat into the hot bath at temperature $T_{ h}$. The resulting heat delivered per cycle into the hot bath, $Q_{ h} = Q_{ c} + W$, consists of the used work and the extracted heat.

In the low-dissipation (LD) regime \cite{PhysRevLett.105.150603,PhysRevA.21.2115,PhysRevE.103.032141}, $Q_i$, $i= c, h$ assume the form
\begin{eqnarray}
Q_{ c}&=&T_{ c}\Delta S-\frac{\sigma_{ c}}{t_{ c}}, 
\label{qc}\\
Q_{ h}&=&T_{ h}\Delta S+\frac{\sigma_{ h}}{t_{ h}}, 
\label{qh}
\end{eqnarray}
where the positive irreversibility parameters $\sigma_i$ depend on the details of system construction, and $t_i$ are durations of the two isotherms. $\Delta S$ denotes the increase (decrease) in the entropy of the system during the cold (hot) isotherm. The corresponding contributions to $Q_c$ and $Q_h$ are reversible, i.e., they do not contribute to the total entropy produced per cycle,
\begin{equation}
\Delta S_{ tot}=-\frac{Q_{ c}}{T_{ c}}+\frac{Q_{ h}}{T_{ h}}=\frac{\sigma_{ c}}{t_{ c}T_{ c}}+\frac{\sigma_{ h}}{t_{ h}T_{ h}}\ge 0.
\label{entropy}
\end{equation}
$\Delta S_{ tot}$ is solely determined by the irreversible contributions, proportional to the irreversibility parameters, and vanishes both in the quasi-static limit, $t_h \to \infty$ and $t_c \to \infty$, and in the quilibrium limit, $\sigma_{ c}=\sigma_{ h}=0$. The LD forms~\eqref{qc} and \eqref{qh} of the transferred heats can be quite generally considered as first-order expansions of the exact expressions in the inverse durations of the isotherms~\cite{Sekimoto1997,Zulkowski2015,Cavina2017,Ma/etal:2020,Holubec2020,PhysRevA.21.2115}. Besides, the LD model is exact for optimized overdamped Brownian heat engines~\cite{schmiedl2007efficiency,Holubec2015} and other specific scenarios \cite{iyyappan2020efficiency,Zulkowski2015}.

We assume that durations of the adiabatic branches are proportional to durations of the isotherms so that the cycle time is given by $t_{ p}=a(t_{h}+t_{c})$. Since the constant $a\ge 1$ only affects the heating load of the pump (see Eq.~\eqref{normalized-power} below), we assume in the rest of the paper that $a=1$. This value corresponds to infinitely fast adiabats \cite{blickle2012realization} and thus maximum heating load as a function of $a$. 

\section{Heating load and COP}
\label{sec-define-variables}

The performance of heat devices is described by their power, $P$,  and efficiency, $\epsilon$. For heat pumps, $P$ and $\epsilon$ are called the coefficient of performance (COP) and the heating load~\cite{ahmadi2015thermo,guo2020equivalent}. $P$ measures the average heat pumped into the hot bath per unit time, and $\epsilon$ shows how much work is needed to pump 1 Joule of heat to the hot body.

Using Eqs. \eqref{qc} and \eqref{qh} together with the first law of thermodynamics, $W=Q_{ h}-Q_{ c}$, the heating load and COP of the LD heat pump can be expressed as
\begin{eqnarray}
P&=&\frac{Q_{ h}}{t_{ p}}=\frac{T_{ h}\Delta S}{t_{ p}}+\frac{\sigma_{ h}}{t_{ h}t_{ p}},
\label{normalized-power}\\
\epsilon&=&\frac{Q_{ h}}{W}=\frac{\epsilon_{ C}}{1+T_{ c}\epsilon_{ C}\Delta S_{ tot}/(Pt_{ p})}.
\label{normalized-effi}
\end{eqnarray}
The maximum (Carnot) COP, $\epsilon=\epsilon_C$, is attained under reversible conditions ($\Delta S_{tot}=0$). The minimum COP, $\epsilon = 1 $, describes the situation when no heat is pumped from the cold bath and thus the delivered heat, $Q_h$, equals the input work, $W$. In this regime, heat pumps are not better than work-to-heat converters, such as resistance heating wires. In the next section, we study COP at maximum heating load for LD heat pumps.

\section{COP at maximum heating load}
\label{sec-four}

Most of the available expressions for maximum efficiency at a fixed power for various models~\cite{Long2016,See-AppendixC,Whitney2014,Whitney2015,PhysRevE.93.050101,holubec2016maximum,Dechant2017,PhysRevE.101.052124,Long2018,ye2020maximum} are given as functions of the dimensionless variable $P/P^*$, measuring loss in power, $P$, with respect to the maximum power, $P^*$. This normalization of power usually significantly simplifies the resulting expressions. However, for endoreversible heat pumps~\cite{leff1978eer,blanchard1980coefficient} the maximum power diverges, suggesting that such a normalization might, in our case, not be possible. Indeed, we show below that $P^*\to\infty$ also for LD heat pumps.

To introduce a meaningful dimensionless heating load, we define the reduced heats and durations as
\begin{equation}
\tilde{Q}_i=\frac{Q_i}{T_{ h}\Delta S},~\tilde{t}_i=\frac{T_{ h}\Delta S}{\sigma_{ h}}t_i,\quad i= c, h.
\label{unit-defined}
\end{equation}
Using Eqs. \eqref{qc} and \eqref{qh}, the reduced heats read
\begin{eqnarray}
\tilde{Q}_{ c}&=&\frac{\epsilon_{ C}-1}{\epsilon_{ C}}-\frac{1}{\sigma(1-\alpha)\tilde{t}_{ p}}.
\label{normalized-qc}\\
\tilde{Q}_{ h}&=&1+\frac{1}{\alpha\tilde{t}_{ p}}.
\label{normalized-qh}
\end{eqnarray}
Here, $\sigma=\sigma_{ h}/\sigma_{ c}$ is the so-called irreversibility ratio,
$\tilde{t}_{p}=\tilde{t}_h+\tilde{t}_c$ denotes the reduced cycle duration, and $\alpha\equiv t_h/t_{ p}$ measures the allocation of the cycle duration between the two isotherms. We define the reduced heating load as the ratio of the reduced heat to the reduced cycle duration:
\begin{equation}
\tilde{P}=\frac{\tilde{Q}_{ h}}{\tilde{t}_{ p}}=\frac{1}{\tilde{t}_{ p}}+\frac{1}{\alpha\tilde{t}_{ p}^2} =
\frac{\sigma_{h}}{(T_{ h}\Delta S)^2} P.
\label{normalized-power-po}
\end{equation}
The reduced heating load is a monotonically decreasing function of both $\alpha$ and $\tilde{t}_{ p}$. The inequality $Q_{h}>Q_{c}>0$, following from the requirement that the considered device pumps heat from the cold to the hot bath, restricts the minimal reduced cycle duration as
\begin{equation}
\tilde{t}_{ p}>\frac{\epsilon_{ C}}{\sigma(\epsilon_{ C}-1)(1-\alpha)}.
\end{equation}
The maximum reduced heating load, $\tilde{P}^*$, attained for the minimal allowed values of $\alpha$ and $\tilde{t}_{ p}$,
\begin{eqnarray}
\alpha^*&=& 0, 
\label{opt-max-lod-alpha}\\
\tilde{t}_{ p}^*&=&\frac{\epsilon_{ C}}{\sigma(\epsilon_{ C}-1)},
\label{opt-max-lod-tp-star}
\end{eqnarray}
hence diverges. The corresponding COP is most easily obtained from the formula $\epsilon= \tilde{Q}_h/(\tilde{Q}_h-\tilde{Q}_c)$. Altogether, the maximum reduced heating load and the corresponding COP read
\begin{eqnarray}
\tilde{P}^*&=&\infty,    
\label{max-heat-load-inf}\\
\epsilon^*&=&1.
\label{max-heat-load-eff-heating}
\end{eqnarray}
This performance is achieved whenever the hot isotherm is much faster than the cold one and thus $\alpha=\alpha^*\to 0$. Noteworthy, the COP at maximum power is the smallest possible, corresponding to the negligible amount of heat pumped from the cold bath compared to the input work, $\tilde{Q}_{ h} = \tilde{W} + \tilde{Q}_{c} \gg \tilde{Q}_{ c}$. A heat pump operating at the maximum heating load thus works as an electric heater transforming work in the form of electric energy into heat. Practical heat pumps should not operate anywhere close to this regime. In the next section, we uncover more practical operation regimes of LD heat pumps by deriving their maximum COP at a given heating load.

\section{Maximum COP at given heating load}
\label{sec-max-COP-at}

Fixing the reduced heating load in Eq.~\eqref{normalized-power-po} creates the dependency
\begin{equation}
\alpha=\frac{1}{\tilde{t}_{ p}(\tilde{P}\tilde{t}_{ p}-1)}
\label{normalized-alpha}
\end{equation}
between $\alpha \in [0,1]$ and $\tilde{t}_p$. Substituting Eq. \eqref{normalized-alpha} into
Eqs.~\eqref{normalized-qc} and \eqref{normalized-qh} and using the condition
$\tilde{Q}_{ h}>\tilde{Q}_{ c}>0$, we find the inequality
\begin{equation}
\tilde{t}_{ p}> \frac{1+\tilde{P}\tilde{t}^*_{ p}}{2\tilde{P}}+\sqrt{\left( \frac{1+\tilde{P}\tilde{t}^*_{ p}}{2\tilde{P}}\right)^2+\frac{1-\tilde{t}^*_{ p}}{\tilde{P}}}\equiv\tilde{t}_{ p, min}.
\label{expand-tmin}
\end{equation}
The minimum value of the reduced cycle duration for fixed heating load, $\tilde{t}_{ p, min}$, thus depends on the irreversibility ratio $\sigma$ and the Carnot COP $\epsilon_{ C}$ via $\tilde{t}_{ p}^*$ in Eq.~\eqref{opt-max-lod-tp-star}. For maximum and minimum values of $\sigma$, $\tilde{t}_{ p, min}$ reads
\begin{equation}
\tilde{t}_{ p, min}=
\begin{cases}
\frac{1+\sqrt{1+4\tilde{P}}}{2\tilde{P}}& \text{for}\quad \sigma\to\infty\\
\infty& \text{for}\quad\sigma\to 0
\end{cases}.    
\label{tilde-p-min-simplify}
\end{equation}
The COP \eqref{normalized-effi} can be written in terms of the reduced parameters introduced above as
\begin{equation}
\epsilon=\left[1+\frac{\tilde{P}\tilde{t}_{ p}-1}{\sigma\tilde{P}\tilde{t}_{p}(\tilde{P}\tilde{t}_{ p}^2-\tilde{t}_{ p}-1)}-\frac{\epsilon_C-1}{\tilde{P}\tilde{t}_{p}\epsilon_C}\right]^{-1}.
\label{COP-t-tmin}
\end{equation}
Below we will find its maximum as a function of $\tilde{t}_{ p}>\tilde{t}_{ p, min}$.

\subsection{Bounds}
\label{sec-bounds}

First, we determine the upper and lower bounds on the maximum COP at a given heating load. Taking the derivative of $\epsilon$ \eqref{COP-t-tmin} with respect to $\sigma$, one finds that $\partial\epsilon/\partial\sigma>0$ and thus $\epsilon$
monotonically increases with $\sigma$. Physically, this is because the COP in Eq.~\eqref{normalized-effi} is for a fixed $P$ and $\sigma_{ h}$ (fixed by our choice of time unit) a monotonically decreasing function of the entropy production, $\Delta S_{tot}$, and thus $\sigma_{ c}$. The lower bound on COP \eqref{COP-t-tmin} for a fixed $P$ is thus attained if the irreversible losses during the hot isotherm are negligible compared to those during the cold one ($\sigma = \sigma_h/\sigma_c \to 0$). The corresponding COP equals 1. Note that due to the condition~\eqref{expand-tmin} the reduced cycle duration $\tilde{t}_{ p}$ in this regime diverges [cf. Eq.~\eqref{tilde-p-min-simplify}]. 

The upper bound on COP \eqref{COP-t-tmin} for a fixed $P$ is attained if irreversible losses during the cold isotherm are negligible compared to those during the hot one ($\sigma\to\infty$). In this regime, the COP,
\begin{equation}
\epsilon=\left(1-\frac{\epsilon_C-1}{\tilde{P}\tilde{t}_{p}\epsilon_C}\right)^{-1},    
\label{limiting-sigma-eff}
\end{equation}
monotonically decreases with $\tilde{t}_{ p}$ and thus it attains its maximum for $\tilde{t}_{ p}=\tilde{t}_{ p, min}=\left(1+\sqrt{1+4\tilde{P}}\right)/(2\tilde{P})$. Altogether, the bounds on the maximum COP at given heating load, $\epsilon^{ opt}=\epsilon^{ opt}(\tilde{P})$, are given by
\begin{equation}
1\le\epsilon^{ opt}\le\frac{(1+\sqrt{1+4\tilde{P}})\epsilon_{ C}}{2-(1-\sqrt{1+4\tilde{P}})\epsilon_{ C}}\equiv\epsilon^{ opt}_>. 
\label{upper-bound-megp}
\end{equation}
As expected, the upper bound, $\epsilon^{ opt}_>$, converges to $\epsilon_{ C}$ for $\tilde{P}\to 0$ and to 1 for $\tilde{P}\to\infty$.

\begin{figure}[t]
\centering
\includegraphics[trim={0cm 0.4cm 0.4cm 0cm}, width=0.9\columnwidth]{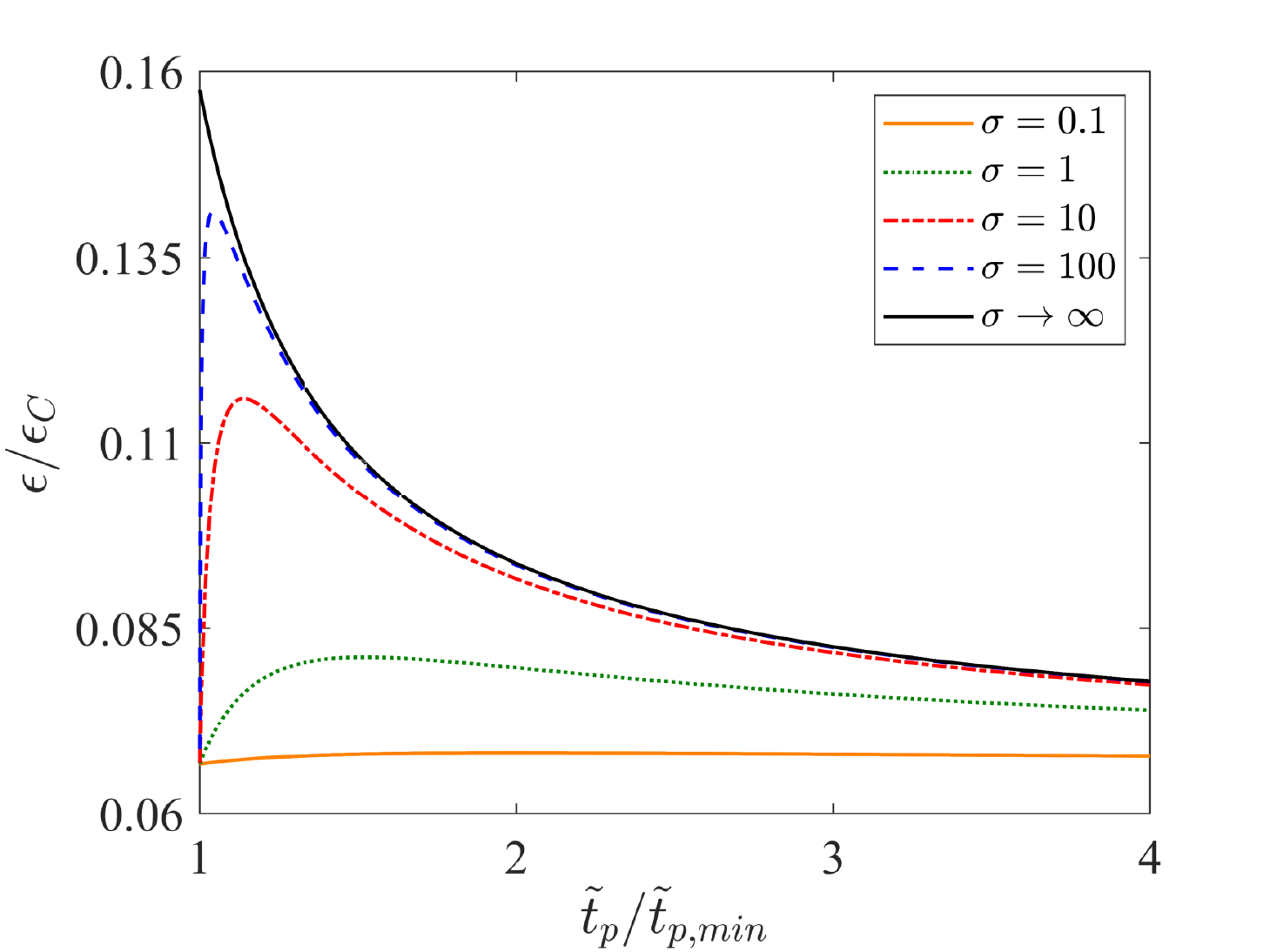}
\caption{COP \eqref{COP-t-tmin} as a function of $\tilde{t}_{ p}/\tilde{t}_{ p, min}$ \eqref{expand-tmin} for different values of $\sigma$, $\tilde{P}=1$, and $\epsilon_{ C}=15$. The figure shows that the upper bound \eqref{upper-bound-megp} on the optimal COP is obtained for $\sigma\to\infty$.}
\label{fig:tp}
\end{figure}

\subsection{Arbitrary parameters}
\label{sec-arbitrary}

Outside the limiting regimes discussed in the previous section, the optimization of COP~\eqref{COP-t-tmin} for a fixed $P$ is more complicated.
In Fig.~\ref{fig:tp}, we show $\epsilon$ as a function of $\tilde{t}_{ p}/\tilde{t}_{ p, min}$ for five values of $\sigma$. The black solid line for $\sigma\to\infty$ indeed monotonously decreases with $\tilde{t}_{ p}$. However, for an arbitrary finite $\sigma$, the COP exhibits a global maximum for $\tilde{t}_{p}^{opt}>\tilde{t}_{ p, min}$. Its position follows from the condition $\partial\epsilon/\partial\tilde{t}_{p}|_{\tilde{t}_{ p}=\tilde{t}_{p}^{ opt}}=0$, which implies the quartic equation
\begin{equation}
\tilde{t}_{p}^4+a\tilde{t}_{p}^3+b\tilde{t}_{p}^2+c\tilde{t}_{p}+\frac{c}{2}=0,  \quad \tilde{t}_{ p}=\tilde{t}_{p}^{ opt},
\label{optmal-formula}
\end{equation}
with the coefficients
\begin{equation}
\left( \begin{array}{c}
a\\
b\\
c
\end{array} \right) 
=\frac{1}{\tilde{P}^2}
\left( \begin{array}{c}
-2\tilde{P}-2\tilde{P}^2\tps\\
1-2\tilde{P}+4\tilde{P}\tps\\
2-2\tps
\end{array} \right).
\end{equation}
Eq.~\eqref{optmal-formula} has four roots which can be determined analytically using
Ferrari's method. The optimal reduced cycle duration is given by the largest real-valued root:
\begin{equation}
\tilde{t}^{opt}_p(\tilde{P}, \tps)=-\frac{a}{4}+ F+\frac{1}{2}\sqrt{-2C-4F^2-\frac{D}{F}},    
\label{optimal-tilde-tp-opt}
\end{equation}
where
\begin{align}
A&=b^2+3c(2-a),\\
B&=2b^3-9bc(4+a)+\frac{27c}{2}(a^2+2c),\\
C&=b-\frac{3a^2}{8},\\
D&=\frac{a^3}{8}-\frac{ab}{2}+c,\\
E&=\sqrt[3]{\frac{B + \sqrt{B^2 - 4A^3}}{2}},\\
F&=\frac{\sqrt{3}}{6}\sqrt{\frac{A}{E}+E-2C}.
\end{align}
For a fixed $\tilde{P}$, the reduced optimal cycle duration only depends on $\tps$ in Eq.~\eqref{opt-max-lod-tp-star}.
Inserting $\tilde{t}^{opt}_p$ into Eq.~\eqref{COP-t-tmin} yields the maximum COP at given heating load for the LD heat pump, $\epsilon^{ opt}=\epsilon^{ opt}(\tilde{P}, \sigma, \epsilon_C)$.

\begin{figure}[t]
\centering
\includegraphics[trim={0.3cm 0.25cm 0cm 0cm}, width=0.85\columnwidth]{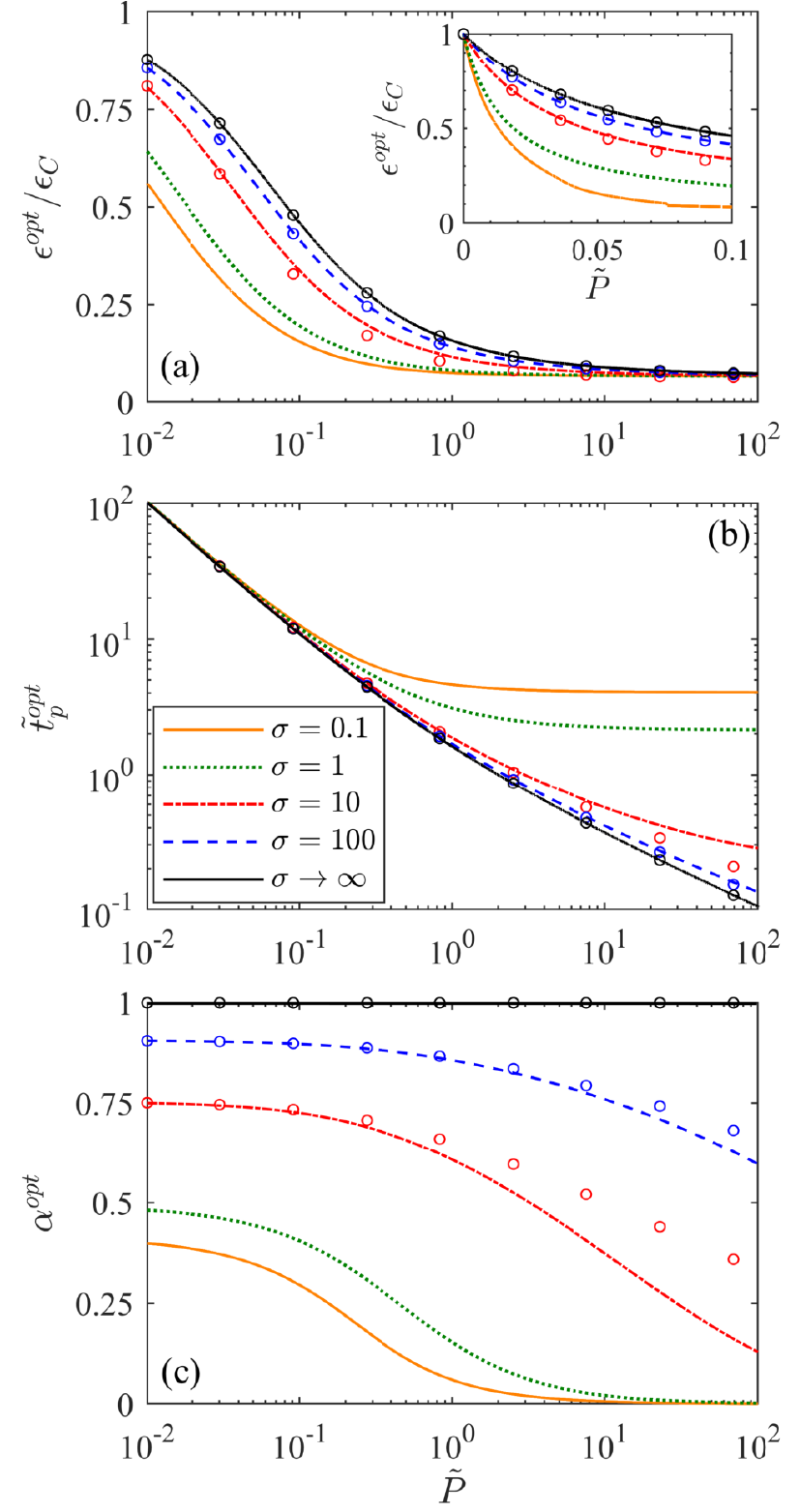}
\caption{The optimal COP (a), the corresponding reduced cycle duration (b), and its allocation between hot and cold isotherms (c) as a function of $\tilde{P}$ for different values of $\sigma$ and $\epsilon_{ C}=15$. The circles follow from the approximate expressions~\eqref{expansion-sigma-infinity} and \eqref{expansion-effi-infinity}. The lines were obtained using the exact result~\eqref{optimal-tilde-tp-opt}.}
\label{fig:parameters}
\end{figure}

In Fig. \ref{fig:parameters}, we show $\epsilon^{opt}$, $\tilde{t}^{opt}_{p}$, and 
$\alpha^{opt} = [\tilde{t}^{opt}_{p}(\tilde{P}\tilde{t}^{opt}_{p}-1)]^{-1}$ [see Eq.~\eqref{normalized-alpha}] as functions of $\tilde{P}$ for five values of $\sigma$. The exact theoretical results are depicted by solid lines. We checked that they agree within numerical precision with the optimal COP obtained by the direct numerical maximization of $\epsilon$ in Eq.~\eqref{COP-t-tmin}. In agreement with the inequalities~\eqref{upper-bound-megp}, $\epsilon^{ opt}$ in Fig.~\ref{fig:parameters}(a) converges to 1 for $\tilde{P}\to \infty$ and to $\epsilon_C$ for $\tilde{P}\to 0$ (see the inset) for all $\sigma$. This panel also shows the monotonic increase of $\epsilon^{ opt}$ with $\sigma$ discussed in Sec.~\ref{sec-bounds}. The increase of the maximum COP with decreasing heating load \textcolor{black}{for large $\tilde{P}$} is very slow, showing that reasonably efficient heat pumps has to operate at small values of $\tilde{P}$. In this respect, heat pumps qualitatively differ from heat engines and refrigerators, which exhibit large gains in efficiency when their power is slightly decreased from its maximum value~\cite{holubec2016maximum,PhysRevE.101.052124}.

The $\sigma$-dependency of $\tilde{t}_{ p}^{ opt}$ in Fig.~\ref{fig:parameters}(b) is significant for small values of $\sigma$ but negligible for large $\sigma$.  Even though the $\sigma$-dependency of $\alpha^{ opt}$ in Fig.~\ref{fig:parameters}(c) is always significant, the COP in Eq.~\eqref{COP-t-tmin} no longer depends on $\alpha$. This suggests that we might obtain an analytically tractable approximation for $\epsilon^{opt}$, valid for intermediate and large values of $\sigma$, by expanding $\tilde{t}_{ p}^{ opt}$ in powers of $\tps \sim 1/\sigma$.
Up to the leading order in $\tps$, we find
\begin{eqnarray}
\tilde{t}_{ p}^{ opt}&\approx&\frac{1+\sqrt{1+4\tilde{P}}}{2\tilde{P}}+\frac{\sqrt{\tilde{t}_{ p}^*}}{(1+4\tilde{P})^{1/4}}, 
\label{expansion-sigma-infinity}  \\
\epsilon^{ opt}&\approx&\epsilon_>^{ opt}-\frac{8\tilde{P}(1-\epsilon_{C}^{-1})(\epsilon_>^{ opt})^2 \sqrt{\tilde{t}_{ p}^*}}{(1+4\tilde{P})^{1/4}(1+\sqrt{1+4\tilde{P}})^2}.
\label{expansion-effi-infinity} 
\end{eqnarray}
The corrections to these formulas are proportional to $\tilde{t}_{p}^*$. For large values of $\tilde{t}_{p}^*$, the approximation \eqref{expansion-effi-infinity} leads to negative (thus unphysical) COP. Circles in Fig.~\ref{fig:parameters} show the predictions from the approximate formulas for $\sigma > 1$, when the approximate $\epsilon^{ opt} > 0$. For large values of $\sigma$ (small $\tilde{t}_{p}^*$), the approximate (circles) and exact (lines) results indeed perfectly overlap. 

\section{Comparison with endoreversible heat pumps} 
\label{comparison-endo}

Let us now compare the obtained results on maximum COP at a given heating load of LD heat pumps to the corresponding known results for endoreversible heat pumps~\cite{blanchard1980coefficient,cheng1995performance,chen1995study}. The endoreversible thermodynamics assumes that the working fluid of thermal devices operates reversibly. The only considered sources of entropy production are the finite-time heat transfers between thermal reservoirs and the working fluid~\cite{Novikov1958,Chambadal1957,curzon1975efficiency}. LD models generally describe the thermodynamics of slowly driven systems~\cite{PhysRevLett.105.150603,PhysRevA.21.2115,PhysRevE.103.032141}. On the other hand, up to a few exceptions~\cite{chen2021microscopic,vaudrey2009detailed,bouton2021quantum},  the endoreversible models are usually phenomenological~\cite{curzon1975efficiency,chen1989effect,rubin1982optimal,chen1989unified,huleihil2006convective}.

The works~\cite{blanchard1980coefficient,cheng1995performance} on the maximum COP at a given heating load of endoreversible heat pumps assume that the heat transfers between the working fluid and baths obey Newton's law of cooling.
Denoting the temperatures of the working fluid during the hot and cold isotherm by $T_{hw}$ and $T_{cw}$ and the corresponding  heat conductivities as $\kappa_h$ and $\kappa_c$, the heats transferred during the isotherms are in this case given by
\begin{eqnarray}
Q_h&=&\kappa_h t_h(T_{hw}-T_h), 
\label{endo-qh}\\
Q_c&=&\kappa_c t_c(T_c-T_{cw}).
\label{endo-qc}
\end{eqnarray}
More general heat transfer laws used in Ref.~\cite{chen1995study} lead to qualitatively the same results as Newton's law of cooling, to which we stick in the following discussion. 

In the endorevesible models, the COP $\varepsilon_{en} = Q_h/(Q_h-Q_c)$ is maximized with respect to the temperatures of the working fluid $T_{hw}$ and $T_{cw}$. The ratio $t_h/t_c$ of the durations of the two isotherms is determined by the endoreversibility requirement $Q_h/T_{hw}- Q_c/T_{cw} = 0$ and the total cycle duration does not influence the resulting expressions. Performing the maximisation for a fixed heating load $P = Q_h/(t_h+t_c)$ with the definitions \eqref{endo-qh} and \eqref{endo-qc} yields the maximum COP~\cite{blanchard1980coefficient,chen1995study}
\begin{equation}
\epsilon^{opt}_{en} = 1+\frac{\epsilon_C-1}{1+\epsilon_C P\left(1+\sqrt{r}\right)^2/(\kappa_h T_h)},
\label{megp-endo}
\end{equation}
\textcolor{black}{where $r=\kappa_h/\kappa_c$.}
The maximum COP at fixed heating load thus behaves qualitatively in the same way as the corresponding result for LD heat pumps: $\epsilon^{opt}_{en}$ converges to $\epsilon_C$
for $P \to 0$ and to $1$ for $P\to \infty$. However, the precise functional forms of the maximum COP for LD and endoreversible heat pumps in general differ. The exception is the parameter regime
\begin{equation}
    \tilde{t}_p^* = 1, \quad 
     \frac{(1+\sqrt{r})^2}{\kappa_h} =  \frac{4\sigma_h}{T_h \Delta S^2},
     \label{eq:PR}
\end{equation}
when the expressions for $\epsilon^{opt}_{en}$ and $\epsilon^{opt}$ are identical. In this regime, one can thus find an exact mapping between the LD and the endoreversible model. Note that for $\tilde{t}_p^* = 1$,
Eq.~\eqref{optmal-formula} reduces to a quadratic equation, and Eq.~\eqref{opt-max-lod-tp-star} implies $\sigma_h/T_h = \sigma_c/T_c$.

\begin{figure}[t]
\centering
\includegraphics[trim={0.35cm 0.55cm 0.6cm 0cm}, width=0.88\columnwidth]{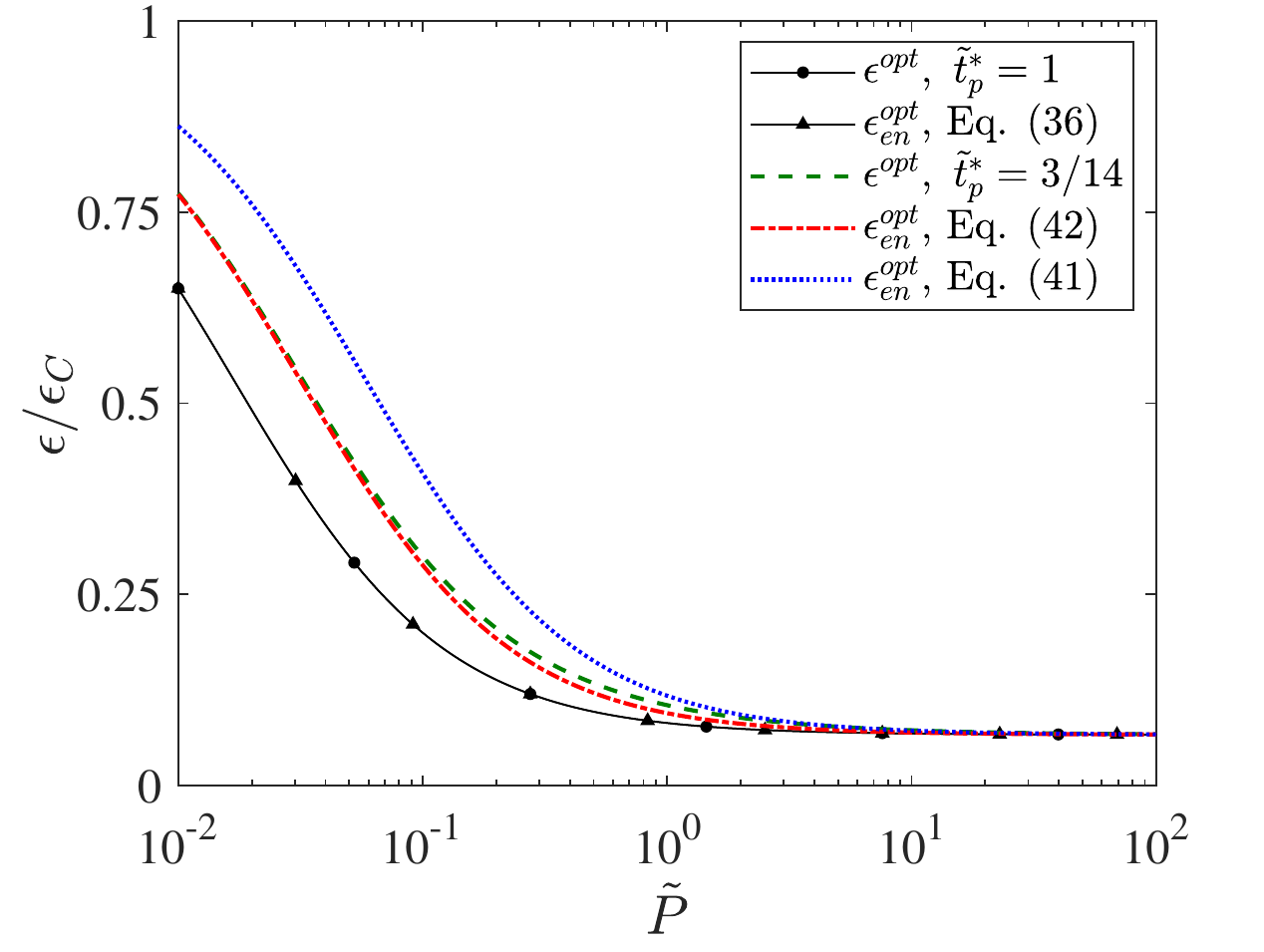}
\caption{The maximum COPs at fixed heating load for LD ($\epsilon^{opt}$) and endoreversible ($\epsilon^{opt}_{en}$) heat pumps as functions of $\tilde{P}$ for  $\epsilon_C=15$ and $\sigma_h/(T_h\Delta S)^2 = 1/(\kappa_h T_h)$. The marked black solid lines show that the condition~\eqref{eq:PR} implies $\epsilon^{opt}=\epsilon^{opt}_{en}$. For the remaining lines, we set $\sigma=5$ and thus $\tilde{t}^*_p = 3/14$. The blue dotted line corresponds to $r$ obtained from Eq.~\eqref{eq:SmP}. For the red dash-dotted line, we calculated $r$ using Eq.~\eqref{eq:LmP}.}
\label{fig:ratio}
\end{figure}

One way to show that the two models are equivalent only in the parameter regime~\eqref{eq:PR} is to compare the formulas for $\epsilon^{opt}_{en}$ and $\epsilon^{opt}$ in the limiting regimes, where they become simple.
To this end, we expand the two maximum COPs as functions of the heating load close to infinite and close to vanishing $P$. Up to the leading order in $P$, the expansions read
\begin{eqnarray}
\epsilon^{opt}&\approx& 1+\frac{1-\epsilon_{C}^{-1}}{4\tps} \frac{(T_h\Delta S)^2}{\sigma_h P},
\label{expand-low-eff-en}\\
\epsilon^{opt}_{en}&\approx& 1+\frac{1-\epsilon_C^{-1}}{\left(1+\sqrt{r}\right)^2} \frac{\kappa_h T_h}{P},    
\label{expand-endo-eff-en}
\end{eqnarray}
and
\begin{eqnarray}
\epsilon^{opt}&\approx& \epsilon_C-\epsilon_C(\epsilon_C-1)\left(1+\sqrt{\tilde{t}_{p}^*}\right)^2 \frac{\sigma_h P}{(T_h\Delta S)^2}
\label{expansion-cop-p-zero}\\
\epsilon^{opt}_{en}&\approx& \epsilon_C-\epsilon_C(\epsilon_C-1)\left(1+\sqrt{r}\right)^2\frac{P}{\kappa_h T_h}.   
\label{expansion-endo-cop-p-zero}
\end{eqnarray}
The corrections to Eqs.~\eqref{expand-low-eff-en} and \eqref{expand-endo-eff-en} are proportional to $1/P^2$ and those to \eqref{expansion-cop-p-zero} and \eqref{expansion-endo-cop-p-zero} are proportional to $P^2$. The LD and endoreversible models for heat pumps can be mapped to each other only if the two types of expansions agree, leading to the conditions
\begin{eqnarray}
    \frac{(1+\sqrt{r})^2}{\kappa_h} &=& \frac{\sigma_h}{T_h \Delta S^2}4\tilde{t}_p^*,
    \label{eq:SmP}\\
    \frac{(1+\sqrt{r})^2}{\kappa_h} &=& \frac{\sigma_h}{T_h \Delta S^2}\left(1+\sqrt{\tilde{t}_p^*}\right)^2.
    \label{eq:LmP}
\end{eqnarray}
The first equality follows from Eqs.~\eqref{expand-low-eff-en} and \eqref{expand-endo-eff-en} and the second one from Eqs.~\eqref{expansion-cop-p-zero} and \eqref{expansion-endo-cop-p-zero}. Requiring validity of both yields the condition~\eqref{eq:PR} \textcolor{black}{(see Appendix \ref{appendix-a-meanging} for more details)}. In Fig. \ref{fig:ratio}, we show $\epsilon^{opt}$ and $\epsilon^{opt}_{en}$ as functions of the reduced heating load $\tilde{P}$. The marked lines show the agreement of $\epsilon^{opt}$ and $\epsilon^{opt}_{en}$ when Eq.~\eqref{eq:PR} holds and thus $\tilde{t}^*_p = 1$.
The remaining lines show $\epsilon^{opt}$ (green dashed line) and $\epsilon^{opt}_{en}$ for parameters obeying solely Eq.~\eqref{eq:SmP} (blue dotted line) and~\eqref{eq:LmP} (red dash-dotted line) for $\tilde{t}^*_p = 3/14$. As expected, the green dashed line only agrees with the blue dotted line for large values of $\tilde{P}$ and with the red dash-dotted line for small values of $\tilde{P}$. 

We tested that also the LD and endoreversible models for heat engines and refrigerators lead to identical results when Eq.~\eqref{eq:PR} holds (data not shown). \textcolor{black}{Besides, an equivalent condition was derived in the linear response regime for heat engines operating at maximum power~\cite{PhysRevE.102.012151,PhysRevE.96.012151}.}

\section{Conclusion and Outlook}
\label{sec-conclusion-outlook}

Like endoreversible heat pumps, Carnot-type low-dissipation (LD) heat pumps operate at maximum power as work to heat converters such as standard electric heaters. Practical heat pumps thus should not operate in this regime. To provide a tool to decide a suitable regime of operation for a given application, we derived an analytical expression for maximum efficiency at a given heating load for LD heat pumps. Besides, we derived upper and lower bounds on this quantity. Qualitatively, our results agree with the corresponding findings obtained earlier for endoreversible heat pumps. Unlike the phenomenological endoreversible models, LD models represent a general first-order finite-time correction to the reversible operation and thus their parameters can be either calculated using a perturbation analysis or measured in experiments. Furthermore, the derived upper bound on the maximum efficiency can be considered as a loose upper bound on the efficiency of heat pumps in general. By adding the result for heat pumps to the known formulas for LD heat engines and refrigerators~\cite{holubec2016maximum,PhysRevE.101.052124,ye2020maximum}, the present paper completes the collection of results for maximum efficiency at a given power for LD thermal devices.

The presented result for maximum efficiency at a given heating load depends on the reduced heating load $\tilde{P}$ in Eq.~\eqref{normalized-power-po}. Therefore, the heating load can be further optimized for the chosen unit of energy flux without affecting the corresponding maximum efficiency. Such optimization tasks performed for LD heat engines and refrigerators are described in Refs.~\cite{PhysRevLett.124.110606, abiuso2020geometric}.
Besides, it would be interesting to investigate the operation regime of maximum efficiency at given power for LD thermal devices concerning its dynamical stability \cite{gonzalez2020energetic,PhysRevE.96.042128,PhysRevE.98.032142,PhysRevE.95.022131}. Finally, it would be worth to investigate maximum efficiency at given power for heat devices operating between finite-sized heat sources \cite{ondrechen1983generalized, izumida2014work,PhysRevE.90.062140,PhysRevE.93.012120,ma2020effect,yuan2021optimizing} and compare the results to those derived using the idealized LD models. \textcolor{black}{For heat engines working with two finite-sized reservoirs, the maximum efficiency at given power has been derived in Ref.~\cite{yuan2021optimizing}.}


\begin{acknowledgments}
ZY is grateful for the sponsorship of China Scholarship Council (CSC) under Grant No. 201906310136. VH gratefully acknowledges support by the Humboldt foundation and by the Czech Science Foundation (project No. 20-02955J).
\end{acknowledgments}

\appendix

\textcolor{black}{
\section{Heat flows in the parameter regime~\eqref{eq:PR}}
\label{appendix-a-meanging}
}

\textcolor{black}{
In this Appendix, we investigate the physical significance of the parameter regime~\eqref{eq:PR}, leading to the same expressions for $\epsilon^{opt}_{en}$ and $\epsilon^{opt}$ for the  endoreversible and LD models. As the average heat flows $Q_h/t_p$ are for the two models fixed to be the same value of power $P$, we focus on the structure of the average heat flows $Q_c/t_p$.
}

\textcolor{black}{
For the endoreversible heat pump, combining Eqs.~\eqref{endo-qh} and~\eqref{endo-qc}, together with the endoreversibility condition $Q^{en}_h/T_{hw}- Q^{en}_c/T_{cw} = 0$ yields (here and below we use the superscript \emph{en} to distinguish between the heats for the endoreversible and LD models) 
\begin{equation}
\frac{Q_c^{en}}{t_p}=\frac{T_cP[P+\kappa_h(T_h-T_{hw})]}{rP(T_h-T_{hw})+T_{hw}[P+\kappa_h(T_h-T_{hw})]}.  
\label{appendix-a-endo-diff}
\end{equation}
For the LD heat pump,
inserting Eq.~\eqref{normalized-alpha} into Eq. \eqref{normalized-qc} implies
\begin{equation}
\frac{Q_c}{t_p}=\frac{T_cT_h\Delta S^2}{\sigma_h}\frac{\tilde{P}\tilde{t}_p^2-\tilde{t}_p(1+\tilde{P}\tps)+\tps-1}{\tilde{t}_p(\tilde{P}\tilde{t}_p^2-\tilde{t}_p-1)}.  
\label{appendix-a-LD-diff}
\end{equation}
Imposing the condition~\eqref{eq:PR} and  returning to dimensional power~\eqref{normalized-power-po}, the heat flow for the LD model changes to 
\begin{equation}
\frac{Q_c}{t_p} =\frac{4 \kappa_h T_c}{\left(\sqrt{r}+1\right)^2}\frac{P \left(\sqrt{r}+1\right)^2 (\tilde{t}_p-1)-4 \kappa_h T_h}{ P \left(\sqrt{r}+1\right)^2 \tilde{t}_p^2-4 \kappa_h T_h (\tilde{t}_p+1)}. 
\label{appendix-a-LD-diff2}
\end{equation}
Interestingly, the functional forms of the heat flows~\eqref{appendix-a-endo-diff} and~\eqref{appendix-a-LD-diff2} in terms of power $P$ and the parameter to be optimized ($T_{hw}$ for the endoreversible and $\tilde{t}_p$ for the LD model) are different, even though the analysis in the main text proves that they must be the same functions of power when $T_{hw}$ and $t_p$ are substituted by the values 
\begin{eqnarray}
T_{hw} &=& T_h + \frac{(1 + \sqrt{r}) P}{\kappa_h},
\\    
\tilde{t}_p &=& 2+\frac{4\kappa_h T_h}{\left(\sqrt{r}+1\right)^2 P},
\end{eqnarray}
maximizing the two heat flows and thus, for fixed power, also the COP~\eqref{normalized-effi}. The formulas for the two heat flows remain different even after the substitutions $T_{hw} = (1 + \sqrt{r}) T/\kappa_h$ and $\tilde{t}_p = 2+ 4\kappa_h T_h/[\left(\sqrt{r}+1\right)^2 T]$, which lead to expressions $\tilde{Q}^{en}_c(T,P)/t_p$ and $\tilde{Q}_c(T,P)/t_p$ exhibiting the same maximum,
\begin{equation}
\frac{\tilde{Q}^{en}_c}{t_p} = \frac{\tilde{Q}_c}{t_p} =\frac{\kappa_hT_cP}{\kappa_h T_h + P(1+\sqrt{r})^2},
\end{equation}
for the same value of $T=P$. The expressions $\tilde{Q}^{en}_c(T,P)/t_p$ and $\tilde{Q}_c(T,P)/t_p$ are thus different unless $T=P$. We conclude that there is no deep physical reason why the performances of the optimized endoreversible and LD models are the same in the parameter regime~\eqref{eq:PR}.}

\bibliography{References}

\begin{thebibliography}{90}%
\makeatletter
\providecommand \@ifxundefined [1]{%
 \@ifx{#1\undefined}
}%
\providecommand \@ifnum [1]{%
 \ifnum #1\expandafter \@firstoftwo
 \else \expandafter \@secondoftwo
 \fi
}%
\providecommand \@ifx [1]{%
 \ifx #1\expandafter \@firstoftwo
 \else \expandafter \@secondoftwo
 \fi
}%
\providecommand \natexlab [1]{#1}%
\providecommand \enquote  [1]{``#1''}%
\providecommand \bibnamefont  [1]{#1}%
\providecommand \bibfnamefont [1]{#1}%
\providecommand \citenamefont [1]{#1}%
\providecommand \href@noop [0]{\@secondoftwo}%
\providecommand \href [0]{\begingroup \@sanitize@url \@href}%
\providecommand \@href[1]{\@@startlink{#1}\@@href}%
\providecommand \@@href[1]{\endgroup#1\@@endlink}%
\providecommand \@sanitize@url [0]{\catcode `\\12\catcode `\$12\catcode
  `\&12\catcode `\#12\catcode `\^12\catcode `\_12\catcode `\%12\relax}%
\providecommand \@@startlink[1]{}%
\providecommand \@@endlink[0]{}%
\providecommand \url  [0]{\begingroup\@sanitize@url \@url }%
\providecommand \@url [1]{\endgroup\@href {#1}{\urlprefix }}%
\providecommand \urlprefix  [0]{URL }%
\providecommand \Eprint [0]{\href }%
\providecommand \doibase [0]{https://doi.org/}%
\providecommand \selectlanguage [0]{\@gobble}%
\providecommand \bibinfo  [0]{\@secondoftwo}%
\providecommand \bibfield  [0]{\@secondoftwo}%
\providecommand \translation [1]{[#1]}%
\providecommand \BibitemOpen [0]{}%
\providecommand \bibitemStop [0]{}%
\providecommand \bibitemNoStop [0]{.\EOS\space}%
\providecommand \EOS [0]{\spacefactor3000\relax}%
\providecommand \BibitemShut  [1]{\csname bibitem#1\endcsname}%
\let\auto@bib@innerbib\@empty
\bibitem [{\citenamefont {Holubec}\ and\ \citenamefont
  {Ryabov}(2015)}]{Holubec2015}%
  \BibitemOpen
  \bibfield  {author} {\bibinfo {author} {\bibfnamefont {V.}~\bibnamefont
  {Holubec}}\ and\ \bibinfo {author} {\bibfnamefont {A.}~\bibnamefont
  {Ryabov}},\ }\bibfield  {title} {\bibinfo {title} {Efficiency at and near
  maximum power of low-dissipation heat engines},\ }\href
  {https://doi.org/10.1103/PhysRevE.92.052125} {\bibfield  {journal} {\bibinfo
  {journal} {Phys. Rev. E}\ }\textbf {\bibinfo {volume} {92}},\ \bibinfo
  {pages} {052125} (\bibinfo {year} {2015})}\BibitemShut {NoStop}%
\bibitem [{\citenamefont {Ammar}\ \emph {et~al.}(2019)\citenamefont {Ammar},
  \citenamefont {Sopian}, \citenamefont {Alghoul}, \citenamefont {Elhub},\ and\
  \citenamefont {Elbreki}}]{ammar2019performance}%
  \BibitemOpen
  \bibfield  {author} {\bibinfo {author} {\bibfnamefont {A.}~\bibnamefont
  {Ammar}}, \bibinfo {author} {\bibfnamefont {K.}~\bibnamefont {Sopian}},
  \bibinfo {author} {\bibfnamefont {M.}~\bibnamefont {Alghoul}}, \bibinfo
  {author} {\bibfnamefont {B.}~\bibnamefont {Elhub}},\ and\ \bibinfo {author}
  {\bibfnamefont {A.}~\bibnamefont {Elbreki}},\ }\bibfield  {title} {\bibinfo
  {title} {Performance study on photovoltaic/thermal solar-assisted heat pump
  system},\ }\href@noop {} {\bibfield  {journal} {\bibinfo  {journal} {J.
  Therm. Anal. Calorim.}\ }\textbf {\bibinfo {volume} {136}},\ \bibinfo {pages}
  {79} (\bibinfo {year} {2019})}\BibitemShut {NoStop}%
\bibitem [{\citenamefont {Chua}\ \emph {et~al.}(2010)\citenamefont {Chua},
  \citenamefont {Chou},\ and\ \citenamefont {Yang}}]{chua2010advances}%
  \BibitemOpen
  \bibfield  {author} {\bibinfo {author} {\bibfnamefont {K.~J.}\ \bibnamefont
  {Chua}}, \bibinfo {author} {\bibfnamefont {S.~K.}\ \bibnamefont {Chou}},\
  and\ \bibinfo {author} {\bibfnamefont {W.}~\bibnamefont {Yang}},\ }\bibfield
  {title} {\bibinfo {title} {Advances in heat pump systems: A review},\
  }\href@noop {} {\bibfield  {journal} {\bibinfo  {journal} {Appl. Energy}\
  }\textbf {\bibinfo {volume} {87}},\ \bibinfo {pages} {3611} (\bibinfo {year}
  {2010})}\BibitemShut {NoStop}%
\bibitem [{\citenamefont {Mohanraj}\ \emph {et~al.}(2021)\citenamefont
  {Mohanraj}, \citenamefont {Karthick},\ and\ \citenamefont
  {Dhivagar}}]{mohanraj2021performance}%
  \BibitemOpen
  \bibfield  {author} {\bibinfo {author} {\bibfnamefont {M.}~\bibnamefont
  {Mohanraj}}, \bibinfo {author} {\bibfnamefont {L.}~\bibnamefont {Karthick}},\
  and\ \bibinfo {author} {\bibfnamefont {R.}~\bibnamefont {Dhivagar}},\
  }\bibfield  {title} {\bibinfo {title} {Performance and economic analysis of a
  heat pump water heater assisted regenerative solar still using latent heat
  storage},\ }\href@noop {} {\bibfield  {journal} {\bibinfo  {journal} {Appl.
  Therm. Eng.}\ }\textbf {\bibinfo {volume} {196}},\ \bibinfo {pages} {117263}
  (\bibinfo {year} {2021})}\BibitemShut {NoStop}%
\bibitem [{\citenamefont {Hu}\ and\ \citenamefont
  {Yuill}(2021)}]{hu2021effects}%
  \BibitemOpen
  \bibfield  {author} {\bibinfo {author} {\bibfnamefont {Y.}~\bibnamefont
  {Hu}}\ and\ \bibinfo {author} {\bibfnamefont {D.~P.}\ \bibnamefont {Yuill}},\
  }\bibfield  {title} {\bibinfo {title} {Effects of multiple simultaneous
  faults on characteristic fault detection features of a heat pump in cooling
  mode},\ }\href@noop {} {\bibfield  {journal} {\bibinfo  {journal} {Energy
  Build.}\ ,\ \bibinfo {pages} {111355}} (\bibinfo {year} {2021})}\BibitemShut
  {NoStop}%
\bibitem [{\citenamefont {Aye}\ \emph {et~al.}(2010)\citenamefont {Aye},
  \citenamefont {Fuller},\ and\ \citenamefont {Canal}}]{aye2010evaluation}%
  \BibitemOpen
  \bibfield  {author} {\bibinfo {author} {\bibfnamefont {L.}~\bibnamefont
  {Aye}}, \bibinfo {author} {\bibfnamefont {R.}~\bibnamefont {Fuller}},\ and\
  \bibinfo {author} {\bibfnamefont {A.}~\bibnamefont {Canal}},\ }\bibfield
  {title} {\bibinfo {title} {Evaluation of a heat pump system for greenhouse
  heating},\ }\href@noop {} {\bibfield  {journal} {\bibinfo  {journal} {Int. J.
  Therm. Sci.}\ }\textbf {\bibinfo {volume} {49}},\ \bibinfo {pages} {202}
  (\bibinfo {year} {2010})}\BibitemShut {NoStop}%
\bibitem [{\citenamefont {Hong}\ and\ \citenamefont
  {Howarth}(2016)}]{hong2016greenhouse}%
  \BibitemOpen
  \bibfield  {author} {\bibinfo {author} {\bibfnamefont {B.}~\bibnamefont
  {Hong}}\ and\ \bibinfo {author} {\bibfnamefont {R.~W.}\ \bibnamefont
  {Howarth}},\ }\bibfield  {title} {\bibinfo {title} {Greenhouse gas emissions
  from domestic hot water: heat pumps compared to most commonly used systems},\
  }\href@noop {} {\bibfield  {journal} {\bibinfo  {journal} {Energy Sci. Eng.}\
  }\textbf {\bibinfo {volume} {4}},\ \bibinfo {pages} {123} (\bibinfo {year}
  {2016})}\BibitemShut {NoStop}%
\bibitem [{\citenamefont {Holubec}\ and\ \citenamefont
  {Ryabov}(2017)}]{PhysRevE.96.062107}%
  \BibitemOpen
  \bibfield  {author} {\bibinfo {author} {\bibfnamefont {V.}~\bibnamefont
  {Holubec}}\ and\ \bibinfo {author} {\bibfnamefont {A.}~\bibnamefont
  {Ryabov}},\ }\bibfield  {title} {\bibinfo {title} {Diverging, but negligible
  power at carnot efficiency: Theory and experiment},\ }\href
  {https://doi.org/10.1103/PhysRevE.96.062107} {\bibfield  {journal} {\bibinfo
  {journal} {Phys. Rev. E}\ }\textbf {\bibinfo {volume} {96}},\ \bibinfo
  {pages} {062107} (\bibinfo {year} {2017})}\BibitemShut {NoStop}%
\bibitem [{\citenamefont {Gordon}\ and\ \citenamefont
  {Orlov}(1993)}]{gordon1993performance}%
  \BibitemOpen
  \bibfield  {author} {\bibinfo {author} {\bibfnamefont {J.}~\bibnamefont
  {Gordon}}\ and\ \bibinfo {author} {\bibfnamefont {V.~N.}\ \bibnamefont
  {Orlov}},\ }\bibfield  {title} {\bibinfo {title} {Performance characteristics
  of endoreversible chemical engines},\ }\href@noop {} {\bibfield  {journal}
  {\bibinfo  {journal} {J. Appl. Phys.}\ }\textbf {\bibinfo {volume} {74}},\
  \bibinfo {pages} {5303} (\bibinfo {year} {1993})}\BibitemShut {NoStop}%
\bibitem [{\citenamefont {Rubin}\ and\ \citenamefont
  {Andresen}(1982)}]{rubin1982optimal}%
  \BibitemOpen
  \bibfield  {author} {\bibinfo {author} {\bibfnamefont {M.~H.}\ \bibnamefont
  {Rubin}}\ and\ \bibinfo {author} {\bibfnamefont {B.}~\bibnamefont
  {Andresen}},\ }\bibfield  {title} {\bibinfo {title} {Optimal staging of
  endoreversible heat engines},\ }\href@noop {} {\bibfield  {journal} {\bibinfo
   {journal} {J. Appl. Phys.}\ }\textbf {\bibinfo {volume} {53}},\ \bibinfo
  {pages} {1} (\bibinfo {year} {1982})}\BibitemShut {NoStop}%
\bibitem [{\citenamefont {Chen}\ and\ \citenamefont
  {Yan}(1989{\natexlab{a}})}]{chen1989unified}%
  \BibitemOpen
  \bibfield  {author} {\bibinfo {author} {\bibfnamefont {J.}~\bibnamefont
  {Chen}}\ and\ \bibinfo {author} {\bibfnamefont {Z.}~\bibnamefont {Yan}},\
  }\bibfield  {title} {\bibinfo {title} {Unified description of endoreversible
  cycles},\ }\href@noop {} {\bibfield  {journal} {\bibinfo  {journal} {Phys.
  Rev. A}\ }\textbf {\bibinfo {volume} {39}},\ \bibinfo {pages} {4140}
  (\bibinfo {year} {1989}{\natexlab{a}})}\BibitemShut {NoStop}%
\bibitem [{\citenamefont {Apertet}\ \emph {et~al.}(2013)\citenamefont
  {Apertet}, \citenamefont {Ouerdane}, \citenamefont {Michot}, \citenamefont
  {Goupil},\ and\ \citenamefont {Lecoeur}}]{apertet2013efficiency}%
  \BibitemOpen
  \bibfield  {author} {\bibinfo {author} {\bibfnamefont {Y.}~\bibnamefont
  {Apertet}}, \bibinfo {author} {\bibfnamefont {H.}~\bibnamefont {Ouerdane}},
  \bibinfo {author} {\bibfnamefont {A.}~\bibnamefont {Michot}}, \bibinfo
  {author} {\bibfnamefont {C.}~\bibnamefont {Goupil}},\ and\ \bibinfo {author}
  {\bibfnamefont {P.}~\bibnamefont {Lecoeur}},\ }\bibfield  {title} {\bibinfo
  {title} {On the efficiency at maximum cooling power},\ }\href@noop {}
  {\bibfield  {journal} {\bibinfo  {journal} {EPL}\ }\textbf {\bibinfo {volume}
  {103}},\ \bibinfo {pages} {40001} (\bibinfo {year} {2013})}\BibitemShut
  {NoStop}%
\bibitem [{\citenamefont {Hu}\ \emph {et~al.}(2013)\citenamefont {Hu},
  \citenamefont {Wu}, \citenamefont {Ma}, \citenamefont {He}, \citenamefont
  {Wang}, \citenamefont {Hern\'andez},\ and\ \citenamefont
  {Roco}}]{PhysRevE.88.062115}%
  \BibitemOpen
  \bibfield  {author} {\bibinfo {author} {\bibfnamefont {Y.}~\bibnamefont
  {Hu}}, \bibinfo {author} {\bibfnamefont {F.}~\bibnamefont {Wu}}, \bibinfo
  {author} {\bibfnamefont {Y.}~\bibnamefont {Ma}}, \bibinfo {author}
  {\bibfnamefont {J.}~\bibnamefont {He}}, \bibinfo {author} {\bibfnamefont
  {J.}~\bibnamefont {Wang}}, \bibinfo {author} {\bibfnamefont {A.~C.}\
  \bibnamefont {Hern\'andez}},\ and\ \bibinfo {author} {\bibfnamefont
  {J.~M.~M.}\ \bibnamefont {Roco}},\ }\bibfield  {title} {\bibinfo {title}
  {Coefficient of performance for a low-dissipation carnot-like refrigerator
  with nonadiabatic dissipation},\ }\href
  {https://doi.org/10.1103/PhysRevE.88.062115} {\bibfield  {journal} {\bibinfo
  {journal} {Phys. Rev. E}\ }\textbf {\bibinfo {volume} {88}},\ \bibinfo
  {pages} {062115} (\bibinfo {year} {2013})}\BibitemShut {NoStop}%
\bibitem [{\citenamefont {Esposito}\ \emph
  {et~al.}(2010{\natexlab{a}})\citenamefont {Esposito}, \citenamefont {Kawai},
  \citenamefont {Lindenberg},\ and\ \citenamefont {Van~den
  Broeck}}]{PhysRevLett.105.150603}%
  \BibitemOpen
  \bibfield  {author} {\bibinfo {author} {\bibfnamefont {M.}~\bibnamefont
  {Esposito}}, \bibinfo {author} {\bibfnamefont {R.}~\bibnamefont {Kawai}},
  \bibinfo {author} {\bibfnamefont {K.}~\bibnamefont {Lindenberg}},\ and\
  \bibinfo {author} {\bibfnamefont {C.}~\bibnamefont {Van~den Broeck}},\
  }\bibfield  {title} {\bibinfo {title} {Efficiency at maximum power of
  low-dissipation carnot engines},\ }\href
  {https://doi.org/10.1103/PhysRevLett.105.150603} {\bibfield  {journal}
  {\bibinfo  {journal} {Phys. Rev. Lett.}\ }\textbf {\bibinfo {volume} {105}},\
  \bibinfo {pages} {150603} (\bibinfo {year} {2010}{\natexlab{a}})}\BibitemShut
  {NoStop}%
\bibitem [{\citenamefont {Gonzalez-Ayala}\ \emph {et~al.}(2020)\citenamefont
  {Gonzalez-Ayala}, \citenamefont {Guo}, \citenamefont {Medina}, \citenamefont
  {Roco},\ and\ \citenamefont {Hern{\'a}ndez}}]{gonzalez2020energetic}%
  \BibitemOpen
  \bibfield  {author} {\bibinfo {author} {\bibfnamefont {J.}~\bibnamefont
  {Gonzalez-Ayala}}, \bibinfo {author} {\bibfnamefont {J.}~\bibnamefont {Guo}},
  \bibinfo {author} {\bibfnamefont {A.}~\bibnamefont {Medina}}, \bibinfo
  {author} {\bibfnamefont {J.}~\bibnamefont {Roco}},\ and\ \bibinfo {author}
  {\bibfnamefont {A.~C.}\ \bibnamefont {Hern{\'a}ndez}},\ }\bibfield  {title}
  {\bibinfo {title} {Energetic self-optimization induced by stability in
  low-dissipation heat engines},\ }\href@noop {} {\bibfield  {journal}
  {\bibinfo  {journal} {Phys. Rev. Lett.}\ }\textbf {\bibinfo {volume} {124}},\
  \bibinfo {pages} {050603} (\bibinfo {year} {2020})}\BibitemShut {NoStop}%
\bibitem [{\citenamefont {Holubec}\ and\ \citenamefont
  {Ye}(2020)}]{PhysRevE.101.052124}%
  \BibitemOpen
  \bibfield  {author} {\bibinfo {author} {\bibfnamefont {V.}~\bibnamefont
  {Holubec}}\ and\ \bibinfo {author} {\bibfnamefont {Z.}~\bibnamefont {Ye}},\
  }\bibfield  {title} {\bibinfo {title} {Maximum efficiency of low-dissipation
  refrigerators at arbitrary cooling power},\ }\href
  {https://doi.org/10.1103/PhysRevE.101.052124} {\bibfield  {journal} {\bibinfo
   {journal} {Phys. Rev. E}\ }\textbf {\bibinfo {volume} {101}},\ \bibinfo
  {pages} {052124} (\bibinfo {year} {2020})}\BibitemShut {NoStop}%
\bibitem [{\citenamefont {Hern{\'a}ndez}\ \emph {et~al.}(2015)\citenamefont
  {Hern{\'a}ndez}, \citenamefont {Medina},\ and\ \citenamefont
  {Roco}}]{hernandez2015time}%
  \BibitemOpen
  \bibfield  {author} {\bibinfo {author} {\bibfnamefont {A.~C.}\ \bibnamefont
  {Hern{\'a}ndez}}, \bibinfo {author} {\bibfnamefont {A.}~\bibnamefont
  {Medina}},\ and\ \bibinfo {author} {\bibfnamefont {J.}~\bibnamefont {Roco}},\
  }\bibfield  {title} {\bibinfo {title} {Time, entropy generation, and
  optimization in low-dissipation heat devices},\ }\href@noop {} {\bibfield
  {journal} {\bibinfo  {journal} {New J. Phys.}\ }\textbf {\bibinfo {volume}
  {17}},\ \bibinfo {pages} {075011} (\bibinfo {year} {2015})}\BibitemShut
  {NoStop}%
\bibitem [{\citenamefont {Benenti}\ \emph {et~al.}(2011)\citenamefont
  {Benenti}, \citenamefont {Saito},\ and\ \citenamefont
  {Casati}}]{benenti2011thermodynamic}%
  \BibitemOpen
  \bibfield  {author} {\bibinfo {author} {\bibfnamefont {G.}~\bibnamefont
  {Benenti}}, \bibinfo {author} {\bibfnamefont {K.}~\bibnamefont {Saito}},\
  and\ \bibinfo {author} {\bibfnamefont {G.}~\bibnamefont {Casati}},\
  }\bibfield  {title} {\bibinfo {title} {Thermodynamic bounds on efficiency for
  systems with broken time-reversal symmetry},\ }\href@noop {} {\bibfield
  {journal} {\bibinfo  {journal} {Phys. Rev. Lett.}\ }\textbf {\bibinfo
  {volume} {106}},\ \bibinfo {pages} {230602} (\bibinfo {year}
  {2011})}\BibitemShut {NoStop}%
\bibitem [{\citenamefont {Izumida}\ and\ \citenamefont
  {Okuda}(2014)}]{izumida2014work}%
  \BibitemOpen
  \bibfield  {author} {\bibinfo {author} {\bibfnamefont {Y.}~\bibnamefont
  {Izumida}}\ and\ \bibinfo {author} {\bibfnamefont {K.}~\bibnamefont
  {Okuda}},\ }\bibfield  {title} {\bibinfo {title} {Work output and efficiency
  at maximum power of linear irreversible heat engines operating with a
  finite-sized heat source},\ }\href@noop {} {\bibfield  {journal} {\bibinfo
  {journal} {Phys. Rev. Lett.}\ }\textbf {\bibinfo {volume} {112}},\ \bibinfo
  {pages} {180603} (\bibinfo {year} {2014})}\BibitemShut {NoStop}%
\bibitem [{\citenamefont {Van~den Broeck}(2005)}]{VandenBroeck2005}%
  \BibitemOpen
  \bibfield  {author} {\bibinfo {author} {\bibfnamefont {C.}~\bibnamefont
  {Van~den Broeck}},\ }\bibfield  {title} {\bibinfo {title} {Thermodynamic
  efficiency at maximum power},\ }\href
  {https://doi.org/10.1103/PhysRevLett.95.190602} {\bibfield  {journal}
  {\bibinfo  {journal} {Phys. Rev. Lett.}\ }\textbf {\bibinfo {volume} {95}},\
  \bibinfo {pages} {190602} (\bibinfo {year} {2005})}\BibitemShut {NoStop}%
\bibitem [{\citenamefont {Izumida}\ \emph {et~al.}(2015)\citenamefont
  {Izumida}, \citenamefont {Okuda}, \citenamefont {Roco},\ and\ \citenamefont
  {Hern{\'a}ndez}}]{Izu2015}%
  \BibitemOpen
  \bibfield  {author} {\bibinfo {author} {\bibfnamefont {Y.}~\bibnamefont
  {Izumida}}, \bibinfo {author} {\bibfnamefont {K.}~\bibnamefont {Okuda}},
  \bibinfo {author} {\bibfnamefont {J.}~\bibnamefont {Roco}},\ and\ \bibinfo
  {author} {\bibfnamefont {A.~C.}\ \bibnamefont {Hern{\'a}ndez}},\ }\bibfield
  {title} {\bibinfo {title} {Heat devices in nonlinear irreversible
  thermodynamics},\ }\href@noop {} {\bibfield  {journal} {\bibinfo  {journal}
  {Phys. Rev. E}\ }\textbf {\bibinfo {volume} {91}},\ \bibinfo {pages} {052140}
  (\bibinfo {year} {2015})}\BibitemShut {NoStop}%
\bibitem [{\citenamefont {Izumida}\ and\ \citenamefont
  {Okuda}(2012)}]{Izu-EPL-heat}%
  \BibitemOpen
  \bibfield  {author} {\bibinfo {author} {\bibfnamefont {Y.}~\bibnamefont
  {Izumida}}\ and\ \bibinfo {author} {\bibfnamefont {K.}~\bibnamefont
  {Okuda}},\ }\bibfield  {title} {\bibinfo {title} {Efficiency at maximum power
  of minimally nonlinear irreversible heat engines},\ }\href@noop {} {\bibfield
   {journal} {\bibinfo  {journal} {EPL}\ }\textbf {\bibinfo {volume} {97}},\
  \bibinfo {pages} {10004} (\bibinfo {year} {2012})}\BibitemShut {NoStop}%
\bibitem [{\citenamefont {Izumida}\ \emph {et~al.}(2013)\citenamefont
  {Izumida}, \citenamefont {Okuda}, \citenamefont {Hern{\'a}ndez},\ and\
  \citenamefont {Roco}}]{Izu-EPL-refri}%
  \BibitemOpen
  \bibfield  {author} {\bibinfo {author} {\bibfnamefont {Y.}~\bibnamefont
  {Izumida}}, \bibinfo {author} {\bibfnamefont {K.}~\bibnamefont {Okuda}},
  \bibinfo {author} {\bibfnamefont {A.~C.}\ \bibnamefont {Hern{\'a}ndez}},\
  and\ \bibinfo {author} {\bibfnamefont {J.}~\bibnamefont {Roco}},\ }\bibfield
  {title} {\bibinfo {title} {Coefficient of performance under optimized figure
  of merit in minimally nonlinear irreversible refrigerator},\ }\href@noop {}
  {\bibfield  {journal} {\bibinfo  {journal} {EPL}\ }\textbf {\bibinfo {volume}
  {101}},\ \bibinfo {pages} {10005} (\bibinfo {year} {2013})}\BibitemShut
  {NoStop}%
\bibitem [{\citenamefont {Uzdin}\ and\ \citenamefont
  {Kosloff}(2014)}]{uzdin2014universal}%
  \BibitemOpen
  \bibfield  {author} {\bibinfo {author} {\bibfnamefont {R.}~\bibnamefont
  {Uzdin}}\ and\ \bibinfo {author} {\bibfnamefont {R.}~\bibnamefont
  {Kosloff}},\ }\bibfield  {title} {\bibinfo {title} {Universal features in the
  efficiency at maximal work of hot quantum otto engines},\ }\href@noop {}
  {\bibfield  {journal} {\bibinfo  {journal} {EPL}\ }\textbf {\bibinfo {volume}
  {108}},\ \bibinfo {pages} {40001} (\bibinfo {year} {2014})}\BibitemShut
  {NoStop}%
\bibitem [{\citenamefont {Abah}\ \emph {et~al.}(2012)\citenamefont {Abah},
  \citenamefont {Rossnagel}, \citenamefont {Jacob}, \citenamefont {Deffner},
  \citenamefont {Schmidt-Kaler}, \citenamefont {Singer},\ and\ \citenamefont
  {Lutz}}]{abah2012single}%
  \BibitemOpen
  \bibfield  {author} {\bibinfo {author} {\bibfnamefont {O.}~\bibnamefont
  {Abah}}, \bibinfo {author} {\bibfnamefont {J.}~\bibnamefont {Rossnagel}},
  \bibinfo {author} {\bibfnamefont {G.}~\bibnamefont {Jacob}}, \bibinfo
  {author} {\bibfnamefont {S.}~\bibnamefont {Deffner}}, \bibinfo {author}
  {\bibfnamefont {F.}~\bibnamefont {Schmidt-Kaler}}, \bibinfo {author}
  {\bibfnamefont {K.}~\bibnamefont {Singer}},\ and\ \bibinfo {author}
  {\bibfnamefont {E.}~\bibnamefont {Lutz}},\ }\bibfield  {title} {\bibinfo
  {title} {Single-ion heat engine at maximum power},\ }\href@noop {} {\bibfield
   {journal} {\bibinfo  {journal} {Phys. Rev. Lett.}\ }\textbf {\bibinfo
  {volume} {109}},\ \bibinfo {pages} {203006} (\bibinfo {year}
  {2012})}\BibitemShut {NoStop}%
\bibitem [{\citenamefont {Ro{\ss}nagel}\ \emph {et~al.}(2014)\citenamefont
  {Ro{\ss}nagel}, \citenamefont {Abah}, \citenamefont {Schmidt-Kaler},
  \citenamefont {Singer},\ and\ \citenamefont {Lutz}}]{rossnagel2014nanoscale}%
  \BibitemOpen
  \bibfield  {author} {\bibinfo {author} {\bibfnamefont {J.}~\bibnamefont
  {Ro{\ss}nagel}}, \bibinfo {author} {\bibfnamefont {O.}~\bibnamefont {Abah}},
  \bibinfo {author} {\bibfnamefont {F.}~\bibnamefont {Schmidt-Kaler}}, \bibinfo
  {author} {\bibfnamefont {K.}~\bibnamefont {Singer}},\ and\ \bibinfo {author}
  {\bibfnamefont {E.}~\bibnamefont {Lutz}},\ }\bibfield  {title} {\bibinfo
  {title} {Nanoscale heat engine beyond the carnot limit},\ }\href@noop {}
  {\bibfield  {journal} {\bibinfo  {journal} {Phys. Rev. Lett.}\ }\textbf
  {\bibinfo {volume} {112}},\ \bibinfo {pages} {030602} (\bibinfo {year}
  {2014})}\BibitemShut {NoStop}%
\bibitem [{\citenamefont {Schmiedl}\ and\ \citenamefont
  {Seifert}(2007)}]{schmiedl2007efficiency}%
  \BibitemOpen
  \bibfield  {author} {\bibinfo {author} {\bibfnamefont {T.}~\bibnamefont
  {Schmiedl}}\ and\ \bibinfo {author} {\bibfnamefont {U.}~\bibnamefont
  {Seifert}},\ }\bibfield  {title} {\bibinfo {title} {Efficiency at maximum
  power: An analytically solvable model for stochastic heat engines},\
  }\href@noop {} {\bibfield  {journal} {\bibinfo  {journal} {EPL}\ }\textbf
  {\bibinfo {volume} {81}},\ \bibinfo {pages} {20003} (\bibinfo {year}
  {2007})}\BibitemShut {NoStop}%
\bibitem [{\citenamefont {Segal}(2008)}]{segal2008stochastic}%
  \BibitemOpen
  \bibfield  {author} {\bibinfo {author} {\bibfnamefont {D.}~\bibnamefont
  {Segal}},\ }\bibfield  {title} {\bibinfo {title} {Stochastic pumping of heat:
  Approaching the carnot efficiency},\ }\href@noop {} {\bibfield  {journal}
  {\bibinfo  {journal} {Phys. Rev. Lett.}\ }\textbf {\bibinfo {volume} {101}},\
  \bibinfo {pages} {260601} (\bibinfo {year} {2008})}\BibitemShut {NoStop}%
\bibitem [{\citenamefont {Jarzynski}\ and\ \citenamefont
  {Mazonka}(1999)}]{jarzynski1999feynman}%
  \BibitemOpen
  \bibfield  {author} {\bibinfo {author} {\bibfnamefont {C.}~\bibnamefont
  {Jarzynski}}\ and\ \bibinfo {author} {\bibfnamefont {O.}~\bibnamefont
  {Mazonka}},\ }\bibfield  {title} {\bibinfo {title} {Feynman’s ratchet and
  pawl: An exactly solvable model},\ }\href@noop {} {\bibfield  {journal}
  {\bibinfo  {journal} {Phys. Rev. E}\ }\textbf {\bibinfo {volume} {59}},\
  \bibinfo {pages} {6448} (\bibinfo {year} {1999})}\BibitemShut {NoStop}%
\bibitem [{\citenamefont {Dechant}\ \emph {et~al.}(2017)\citenamefont
  {Dechant}, \citenamefont {Kiesel},\ and\ \citenamefont {Lutz}}]{Dechant2017}%
  \BibitemOpen
  \bibfield  {author} {\bibinfo {author} {\bibfnamefont {A.}~\bibnamefont
  {Dechant}}, \bibinfo {author} {\bibfnamefont {N.}~\bibnamefont {Kiesel}},\
  and\ \bibinfo {author} {\bibfnamefont {E.}~\bibnamefont {Lutz}},\ }\bibfield
  {title} {\bibinfo {title} {Underdamped stochastic heat engine at maximum
  efficiency},\ }\href {https://doi.org/10.1209/0295-5075/119/50003} {\bibfield
   {journal} {\bibinfo  {journal} {EPL}\ }\textbf {\bibinfo {volume} {119}},\
  \bibinfo {pages} {50003} (\bibinfo {year} {2017})}\BibitemShut {NoStop}%
\bibitem [{\citenamefont {Holubec}(2014)}]{Holubec2014}%
  \BibitemOpen
  \bibfield  {author} {\bibinfo {author} {\bibfnamefont {V.}~\bibnamefont
  {Holubec}},\ }\bibfield  {title} {\bibinfo {title} {An exactly solvable model
  of a stochastic heat engine: optimization of power, power fluctuations and
  efficiency},\ }\href {https://doi.org/10.1088/1742-5468/2014/05/p05022}
  {\bibfield  {journal} {\bibinfo  {journal} {J. Stat. Mech.}\ }\textbf
  {\bibinfo {volume} {2014}},\ \bibinfo {pages} {P05022} (\bibinfo {year}
  {2014})}\BibitemShut {NoStop}%
\bibitem [{\citenamefont {Esposito}\ \emph {et~al.}(2009)\citenamefont
  {Esposito}, \citenamefont {Lindenberg},\ and\ \citenamefont {Van~den
  Broeck}}]{PhysRevLett.102.130602}%
  \BibitemOpen
  \bibfield  {author} {\bibinfo {author} {\bibfnamefont {M.}~\bibnamefont
  {Esposito}}, \bibinfo {author} {\bibfnamefont {K.}~\bibnamefont
  {Lindenberg}},\ and\ \bibinfo {author} {\bibfnamefont {C.}~\bibnamefont
  {Van~den Broeck}},\ }\bibfield  {title} {\bibinfo {title} {Universality of
  efficiency at maximum power},\ }\href
  {https://doi.org/10.1103/PhysRevLett.102.130602} {\bibfield  {journal}
  {\bibinfo  {journal} {Phys. Rev. Lett.}\ }\textbf {\bibinfo {volume} {102}},\
  \bibinfo {pages} {130602} (\bibinfo {year} {2009})}\BibitemShut {NoStop}%
\bibitem [{\citenamefont {Sheng}\ and\ \citenamefont
  {Tu}(2014)}]{PhysRevE.89.012129}%
  \BibitemOpen
  \bibfield  {author} {\bibinfo {author} {\bibfnamefont {S.}~\bibnamefont
  {Sheng}}\ and\ \bibinfo {author} {\bibfnamefont {Z.~C.}\ \bibnamefont {Tu}},\
  }\bibfield  {title} {\bibinfo {title} {Weighted reciprocal of temperature,
  weighted thermal flux, and their applications in finite-time
  thermodynamics},\ }\href {https://doi.org/10.1103/PhysRevE.89.012129}
  {\bibfield  {journal} {\bibinfo  {journal} {Phys. Rev. E}\ }\textbf {\bibinfo
  {volume} {89}},\ \bibinfo {pages} {012129} (\bibinfo {year}
  {2014})}\BibitemShut {NoStop}%
\bibitem [{\citenamefont {Sheng}\ and\ \citenamefont
  {Tu}(2015)}]{PhysRevE.91.022136}%
  \BibitemOpen
  \bibfield  {author} {\bibinfo {author} {\bibfnamefont {S.}~\bibnamefont
  {Sheng}}\ and\ \bibinfo {author} {\bibfnamefont {Z.~C.}\ \bibnamefont {Tu}},\
  }\bibfield  {title} {\bibinfo {title} {Constitutive relation for nonlinear
  response and universality of efficiency at maximum power for tight-coupling
  heat engines},\ }\href {https://doi.org/10.1103/PhysRevE.91.022136}
  {\bibfield  {journal} {\bibinfo  {journal} {Phys. Rev. E}\ }\textbf {\bibinfo
  {volume} {91}},\ \bibinfo {pages} {022136} (\bibinfo {year}
  {2015})}\BibitemShut {NoStop}%
\bibitem [{\citenamefont {Ye}\ \emph {et~al.}(2017)\citenamefont {Ye},
  \citenamefont {Hu}, \citenamefont {He},\ and\ \citenamefont
  {Wang}}]{ye2017universality}%
  \BibitemOpen
  \bibfield  {author} {\bibinfo {author} {\bibfnamefont {Z.}~\bibnamefont
  {Ye}}, \bibinfo {author} {\bibfnamefont {Y.}~\bibnamefont {Hu}}, \bibinfo
  {author} {\bibfnamefont {J.}~\bibnamefont {He}},\ and\ \bibinfo {author}
  {\bibfnamefont {J.}~\bibnamefont {Wang}},\ }\bibfield  {title} {\bibinfo
  {title} {Universality of maximum-work efficiency of a cyclic heat engine
  based on a finite system of ultracold atoms},\ }\href@noop {} {\bibfield
  {journal} {\bibinfo  {journal} {Sci. Rep.}\ }\textbf {\bibinfo {volume}
  {7}},\ \bibinfo {pages} {1} (\bibinfo {year} {2017})}\BibitemShut {NoStop}%
\bibitem [{\citenamefont {Guo}\ \emph {et~al.}(2013)\citenamefont {Guo},
  \citenamefont {Wang}, \citenamefont {Wang},\ and\ \citenamefont
  {Chen}}]{PhysRevE.87.012133}%
  \BibitemOpen
  \bibfield  {author} {\bibinfo {author} {\bibfnamefont {J.}~\bibnamefont
  {Guo}}, \bibinfo {author} {\bibfnamefont {J.}~\bibnamefont {Wang}}, \bibinfo
  {author} {\bibfnamefont {Y.}~\bibnamefont {Wang}},\ and\ \bibinfo {author}
  {\bibfnamefont {J.}~\bibnamefont {Chen}},\ }\bibfield  {title} {\bibinfo
  {title} {Universal efficiency bounds of weak-dissipative thermodynamic cycles
  at the maximum power output},\ }\href
  {https://doi.org/10.1103/PhysRevE.87.012133} {\bibfield  {journal} {\bibinfo
  {journal} {Phys. Rev. E}\ }\textbf {\bibinfo {volume} {87}},\ \bibinfo
  {pages} {012133} (\bibinfo {year} {2013})}\BibitemShut {NoStop}%
\bibitem [{\citenamefont {Esposito}\ \emph
  {et~al.}(2010{\natexlab{b}})\citenamefont {Esposito}, \citenamefont {Kawai},
  \citenamefont {Lindenberg},\ and\ \citenamefont {Van~den
  Broeck}}]{PhysRevE.81.041106}%
  \BibitemOpen
  \bibfield  {author} {\bibinfo {author} {\bibfnamefont {M.}~\bibnamefont
  {Esposito}}, \bibinfo {author} {\bibfnamefont {R.}~\bibnamefont {Kawai}},
  \bibinfo {author} {\bibfnamefont {K.}~\bibnamefont {Lindenberg}},\ and\
  \bibinfo {author} {\bibfnamefont {C.}~\bibnamefont {Van~den Broeck}},\
  }\bibfield  {title} {\bibinfo {title} {Quantum-dot carnot engine at maximum
  power},\ }\href {https://doi.org/10.1103/PhysRevE.81.041106} {\bibfield
  {journal} {\bibinfo  {journal} {Phys. Rev. E}\ }\textbf {\bibinfo {volume}
  {81}},\ \bibinfo {pages} {041106} (\bibinfo {year}
  {2010}{\natexlab{b}})}\BibitemShut {NoStop}%
\bibitem [{\citenamefont {Allahverdyan}\ \emph {et~al.}(2008)\citenamefont
  {Allahverdyan}, \citenamefont {Johal},\ and\ \citenamefont
  {Mahler}}]{PhysRevE.77.041118}%
  \BibitemOpen
  \bibfield  {author} {\bibinfo {author} {\bibfnamefont {A.~E.}\ \bibnamefont
  {Allahverdyan}}, \bibinfo {author} {\bibfnamefont {R.~S.}\ \bibnamefont
  {Johal}},\ and\ \bibinfo {author} {\bibfnamefont {G.}~\bibnamefont
  {Mahler}},\ }\bibfield  {title} {\bibinfo {title} {Work extremum principle:
  Structure and function of quantum heat engines},\ }\href
  {https://doi.org/10.1103/PhysRevE.77.041118} {\bibfield  {journal} {\bibinfo
  {journal} {Phys. Rev. E}\ }\textbf {\bibinfo {volume} {77}},\ \bibinfo
  {pages} {041118} (\bibinfo {year} {2008})}\BibitemShut {NoStop}%
\bibitem [{\citenamefont {Allahverdyan}\ \emph {et~al.}(2013)\citenamefont
  {Allahverdyan}, \citenamefont {Hovhannisyan}, \citenamefont {Melkikh},\ and\
  \citenamefont {Gevorkian}}]{PhysRevLett.111.050601}%
  \BibitemOpen
  \bibfield  {author} {\bibinfo {author} {\bibfnamefont {A.~E.}\ \bibnamefont
  {Allahverdyan}}, \bibinfo {author} {\bibfnamefont {K.~V.}\ \bibnamefont
  {Hovhannisyan}}, \bibinfo {author} {\bibfnamefont {A.~V.}\ \bibnamefont
  {Melkikh}},\ and\ \bibinfo {author} {\bibfnamefont {S.~G.}\ \bibnamefont
  {Gevorkian}},\ }\bibfield  {title} {\bibinfo {title} {Carnot cycle at finite
  power: Attainability of maximal efficiency},\ }\href
  {https://doi.org/10.1103/PhysRevLett.111.050601} {\bibfield  {journal}
  {\bibinfo  {journal} {Phys. Rev. Lett.}\ }\textbf {\bibinfo {volume} {111}},\
  \bibinfo {pages} {050601} (\bibinfo {year} {2013})}\BibitemShut {NoStop}%
\bibitem [{\citenamefont {Long}\ \emph {et~al.}(2018)\citenamefont {Long},
  \citenamefont {Liu},\ and\ \citenamefont {Liu}}]{Long2018}%
  \BibitemOpen
  \bibfield  {author} {\bibinfo {author} {\bibfnamefont {R.}~\bibnamefont
  {Long}}, \bibinfo {author} {\bibfnamefont {Z.}~\bibnamefont {Liu}},\ and\
  \bibinfo {author} {\bibfnamefont {W.}~\bibnamefont {Liu}},\ }\bibfield
  {title} {\bibinfo {title} {Performance analysis for minimally nonlinear
  irreversible refrigerators at finite cooling power},\ }\href@noop {}
  {\bibfield  {journal} {\bibinfo  {journal} {Phys. A: Stat. Mech. Appl.}\
  }\textbf {\bibinfo {volume} {496}},\ \bibinfo {pages} {137} (\bibinfo {year}
  {2018})}\BibitemShut {NoStop}%
\bibitem [{\citenamefont {Leff}\ and\ \citenamefont
  {Teeters}(1978)}]{leff1978eer}%
  \BibitemOpen
  \bibfield  {author} {\bibinfo {author} {\bibfnamefont {H.~S.}\ \bibnamefont
  {Leff}}\ and\ \bibinfo {author} {\bibfnamefont {W.~D.}\ \bibnamefont
  {Teeters}},\ }\bibfield  {title} {\bibinfo {title} {Eer, cop, and the second
  law efficiency for air conditioners},\ }\href@noop {} {\bibfield  {journal}
  {\bibinfo  {journal} {Am. J. Phys.}\ }\textbf {\bibinfo {volume} {46}},\
  \bibinfo {pages} {19} (\bibinfo {year} {1978})}\BibitemShut {NoStop}%
\bibitem [{\citenamefont {Blanchard}(1980)}]{blanchard1980coefficient}%
  \BibitemOpen
  \bibfield  {author} {\bibinfo {author} {\bibfnamefont {C.}~\bibnamefont
  {Blanchard}},\ }\bibfield  {title} {\bibinfo {title} {Coefficient of
  performance for finite speed heat pump},\ }\href@noop {} {\bibfield
  {journal} {\bibinfo  {journal} {J. Appl. Phys.}\ }\textbf {\bibinfo {volume}
  {51}},\ \bibinfo {pages} {2471} (\bibinfo {year} {1980})}\BibitemShut
  {NoStop}%
\bibitem [{\citenamefont {Yan}\ and\ \citenamefont
  {Chen}(1990)}]{yan1990class}%
  \BibitemOpen
  \bibfield  {author} {\bibinfo {author} {\bibfnamefont {Z.}~\bibnamefont
  {Yan}}\ and\ \bibinfo {author} {\bibfnamefont {J.}~\bibnamefont {Chen}},\
  }\bibfield  {title} {\bibinfo {title} {A class of irreversible carnot
  refrigeration cycles with a general heat transfer law},\ }\href@noop {}
  {\bibfield  {journal} {\bibinfo  {journal} {J. Phys. D}\ }\textbf {\bibinfo
  {volume} {23}},\ \bibinfo {pages} {136} (\bibinfo {year} {1990})}\BibitemShut
  {NoStop}%
\bibitem [{\citenamefont {de~Tom{\'a}s}\ \emph {et~al.}(2012)\citenamefont
  {de~Tom{\'a}s}, \citenamefont {Hern{\'a}ndez},\ and\ \citenamefont
  {Roco}}]{de2012optimal}%
  \BibitemOpen
  \bibfield  {author} {\bibinfo {author} {\bibfnamefont {C.}~\bibnamefont
  {de~Tom{\'a}s}}, \bibinfo {author} {\bibfnamefont {A.~C.}\ \bibnamefont
  {Hern{\'a}ndez}},\ and\ \bibinfo {author} {\bibfnamefont {J.}~\bibnamefont
  {Roco}},\ }\bibfield  {title} {\bibinfo {title} {Optimal low symmetric
  dissipation carnot engines and refrigerators},\ }\href@noop {} {\bibfield
  {journal} {\bibinfo  {journal} {Phys. Rev. E}\ }\textbf {\bibinfo {volume}
  {85}},\ \bibinfo {pages} {010104(R)} (\bibinfo {year} {2012})}\BibitemShut
  {NoStop}%
\bibitem [{\citenamefont {Hern{\'a}ndez}\ \emph {et~al.}(2001)\citenamefont
  {Hern{\'a}ndez}, \citenamefont {Medina}, \citenamefont {Roco}, \citenamefont
  {White},\ and\ \citenamefont {Velasco}}]{hernandez2001unified}%
  \BibitemOpen
  \bibfield  {author} {\bibinfo {author} {\bibfnamefont {A.~C.}\ \bibnamefont
  {Hern{\'a}ndez}}, \bibinfo {author} {\bibfnamefont {A.}~\bibnamefont
  {Medina}}, \bibinfo {author} {\bibfnamefont {J.}~\bibnamefont {Roco}},
  \bibinfo {author} {\bibfnamefont {J.}~\bibnamefont {White}},\ and\ \bibinfo
  {author} {\bibfnamefont {S.}~\bibnamefont {Velasco}},\ }\bibfield  {title}
  {\bibinfo {title} {Unified optimization criterion for energy converters},\
  }\href@noop {} {\bibfield  {journal} {\bibinfo  {journal} {Phys. Rev. E}\
  }\textbf {\bibinfo {volume} {63}},\ \bibinfo {pages} {037102} (\bibinfo
  {year} {2001})}\BibitemShut {NoStop}%
\bibitem [{\citenamefont {Angulo-Brown}(1991)}]{angulo1991ecological}%
  \BibitemOpen
  \bibfield  {author} {\bibinfo {author} {\bibfnamefont {F.}~\bibnamefont
  {Angulo-Brown}},\ }\bibfield  {title} {\bibinfo {title} {An ecological
  optimization criterion for finite-time heat engines},\ }\href@noop {}
  {\bibfield  {journal} {\bibinfo  {journal} {J. Appl. Phys.}\ }\textbf
  {\bibinfo {volume} {69}},\ \bibinfo {pages} {7465} (\bibinfo {year}
  {1991})}\BibitemShut {NoStop}%
\bibitem [{\citenamefont {De~Tom{\'a}s}\ \emph {et~al.}(2013)\citenamefont
  {De~Tom{\'a}s}, \citenamefont {Roco}, \citenamefont {Hern{\'a}ndez},
  \citenamefont {Wang},\ and\ \citenamefont {Tu}}]{de2013low}%
  \BibitemOpen
  \bibfield  {author} {\bibinfo {author} {\bibfnamefont {C.}~\bibnamefont
  {De~Tom{\'a}s}}, \bibinfo {author} {\bibfnamefont {J.}~\bibnamefont {Roco}},
  \bibinfo {author} {\bibfnamefont {A.~C.}\ \bibnamefont {Hern{\'a}ndez}},
  \bibinfo {author} {\bibfnamefont {Y.}~\bibnamefont {Wang}},\ and\ \bibinfo
  {author} {\bibfnamefont {Z.}~\bibnamefont {Tu}},\ }\bibfield  {title}
  {\bibinfo {title} {Low-dissipation heat devices: Unified trade-off
  optimization and bounds},\ }\href@noop {} {\bibfield  {journal} {\bibinfo
  {journal} {Phys. Rev. E}\ }\textbf {\bibinfo {volume} {87}},\ \bibinfo
  {pages} {012105} (\bibinfo {year} {2013})}\BibitemShut {NoStop}%
\bibitem [{\citenamefont {S\'anchez-Salas}\ \emph {et~al.}(2010)\citenamefont
  {S\'anchez-Salas}, \citenamefont {L\'opez-Palacios}, \citenamefont
  {Velasco},\ and\ \citenamefont {Calvo~Hern\'andez}}]{PhysRevE.82.051101}%
  \BibitemOpen
  \bibfield  {author} {\bibinfo {author} {\bibfnamefont {N.}~\bibnamefont
  {S\'anchez-Salas}}, \bibinfo {author} {\bibfnamefont {L.}~\bibnamefont
  {L\'opez-Palacios}}, \bibinfo {author} {\bibfnamefont {S.}~\bibnamefont
  {Velasco}},\ and\ \bibinfo {author} {\bibfnamefont {A.}~\bibnamefont
  {Calvo~Hern\'andez}},\ }\bibfield  {title} {\bibinfo {title} {Optimization
  criteria, bounds, and efficiencies of heat engines},\ }\href
  {https://doi.org/10.1103/PhysRevE.82.051101} {\bibfield  {journal} {\bibinfo
  {journal} {Phys. Rev. E}\ }\textbf {\bibinfo {volume} {82}},\ \bibinfo
  {pages} {051101} (\bibinfo {year} {2010})}\BibitemShut {NoStop}%
\bibitem [{\citenamefont {Zhang}\ \emph {et~al.}(2016)\citenamefont {Zhang},
  \citenamefont {Huang}, \citenamefont {Lin},\ and\ \citenamefont
  {Chen}}]{PhysRevE.93.032152}%
  \BibitemOpen
  \bibfield  {author} {\bibinfo {author} {\bibfnamefont {Y.}~\bibnamefont
  {Zhang}}, \bibinfo {author} {\bibfnamefont {C.}~\bibnamefont {Huang}},
  \bibinfo {author} {\bibfnamefont {G.}~\bibnamefont {Lin}},\ and\ \bibinfo
  {author} {\bibfnamefont {J.}~\bibnamefont {Chen}},\ }\bibfield  {title}
  {\bibinfo {title} {Universality of efficiency at unified trade-off
  optimization},\ }\href {https://doi.org/10.1103/PhysRevE.93.032152}
  {\bibfield  {journal} {\bibinfo  {journal} {Phys. Rev. E}\ }\textbf {\bibinfo
  {volume} {93}},\ \bibinfo {pages} {032152} (\bibinfo {year}
  {2016})}\BibitemShut {NoStop}%
\bibitem [{\citenamefont {Long}\ and\ \citenamefont
  {Liu}(2015)}]{long2015ecological}%
  \BibitemOpen
  \bibfield  {author} {\bibinfo {author} {\bibfnamefont {R.}~\bibnamefont
  {Long}}\ and\ \bibinfo {author} {\bibfnamefont {W.}~\bibnamefont {Liu}},\
  }\bibfield  {title} {\bibinfo {title} {Ecological optimization for general
  heat engines},\ }\href@noop {} {\bibfield  {journal} {\bibinfo  {journal}
  {Phys. A: Stat. Mech. Appl.}\ }\textbf {\bibinfo {volume} {434}},\ \bibinfo
  {pages} {232} (\bibinfo {year} {2015})}\BibitemShut {NoStop}%
\bibitem [{\citenamefont {Ryabov}\ and\ \citenamefont
  {Holubec}(2016)}]{PhysRevE.93.050101}%
  \BibitemOpen
  \bibfield  {author} {\bibinfo {author} {\bibfnamefont {A.}~\bibnamefont
  {Ryabov}}\ and\ \bibinfo {author} {\bibfnamefont {V.}~\bibnamefont
  {Holubec}},\ }\bibfield  {title} {\bibinfo {title} {Maximum efficiency of
  steady-state heat engines at arbitrary power},\ }\href
  {https://doi.org/10.1103/PhysRevE.93.050101} {\bibfield  {journal} {\bibinfo
  {journal} {Phys. Rev. E}\ }\textbf {\bibinfo {volume} {93}},\ \bibinfo
  {pages} {050101} (\bibinfo {year} {2016})}\BibitemShut {NoStop}%
\bibitem [{\citenamefont {Johal}(2017)}]{PhysRevE.96.012151}%
  \BibitemOpen
  \bibfield  {author} {\bibinfo {author} {\bibfnamefont {R.~S.}\ \bibnamefont
  {Johal}},\ }\bibfield  {title} {\bibinfo {title} {Heat engines at optimal
  power: Low-dissipation versus endoreversible model},\ }\href
  {https://doi.org/10.1103/PhysRevE.96.012151} {\bibfield  {journal} {\bibinfo
  {journal} {Phys. Rev. E}\ }\textbf {\bibinfo {volume} {96}},\ \bibinfo
  {pages} {012151} (\bibinfo {year} {2017})}\BibitemShut {NoStop}%
\bibitem [{\citenamefont {Zhang}\ and\ \citenamefont
  {Huang}(2020)}]{PhysRevE.102.012151}%
  \BibitemOpen
  \bibfield  {author} {\bibinfo {author} {\bibfnamefont {Y.}~\bibnamefont
  {Zhang}}\ and\ \bibinfo {author} {\bibfnamefont {Y.}~\bibnamefont {Huang}},\
  }\bibfield  {title} {\bibinfo {title} {Applicability of the low-dissipation
  model: Carnot-like heat engines under newton's law of cooling},\ }\href
  {https://doi.org/10.1103/PhysRevE.102.012151} {\bibfield  {journal} {\bibinfo
   {journal} {Phys. Rev. E}\ }\textbf {\bibinfo {volume} {102}},\ \bibinfo
  {pages} {012151} (\bibinfo {year} {2020})}\BibitemShut {NoStop}%
\bibitem [{\citenamefont {Ma}\ \emph {et~al.}(2018)\citenamefont {Ma},
  \citenamefont {Xu}, \citenamefont {Dong},\ and\ \citenamefont
  {Sun}}]{See-AppendixC}%
  \BibitemOpen
  \bibfield  {author} {\bibinfo {author} {\bibfnamefont {Y.-H.}\ \bibnamefont
  {Ma}}, \bibinfo {author} {\bibfnamefont {D.}~\bibnamefont {Xu}}, \bibinfo
  {author} {\bibfnamefont {H.}~\bibnamefont {Dong}},\ and\ \bibinfo {author}
  {\bibfnamefont {C.-P.}\ \bibnamefont {Sun}},\ }\bibfield  {title} {\bibinfo
  {title} {Universal constraint for efficiency and power of a low-dissipation
  heat engine},\ }\href@noop {} {\bibfield  {journal} {\bibinfo  {journal}
  {Phys. Rev. E}\ }\textbf {\bibinfo {volume} {98}},\ \bibinfo {pages} {042112}
  (\bibinfo {year} {2018})}\BibitemShut {NoStop}%
\bibitem [{\citenamefont {Whitney}(2014)}]{Whitney2014}%
  \BibitemOpen
  \bibfield  {author} {\bibinfo {author} {\bibfnamefont {R.~S.}\ \bibnamefont
  {Whitney}},\ }\bibfield  {title} {\bibinfo {title} {Most efficient quantum
  thermoelectric at finite power output},\ }\href
  {https://doi.org/10.1103/PhysRevLett.112.130601} {\bibfield  {journal}
  {\bibinfo  {journal} {Phys. Rev. Lett.}\ }\textbf {\bibinfo {volume} {112}},\
  \bibinfo {pages} {130601} (\bibinfo {year} {2014})}\BibitemShut {NoStop}%
\bibitem [{\citenamefont {Whitney}(2015)}]{Whitney2015}%
  \BibitemOpen
  \bibfield  {author} {\bibinfo {author} {\bibfnamefont {R.~S.}\ \bibnamefont
  {Whitney}},\ }\bibfield  {title} {\bibinfo {title} {Finding the quantum
  thermoelectric with maximal efficiency and minimal entropy production at
  given power output},\ }\href {https://doi.org/10.1103/PhysRevB.91.115425}
  {\bibfield  {journal} {\bibinfo  {journal} {Phys. Rev. B}\ }\textbf {\bibinfo
  {volume} {91}},\ \bibinfo {pages} {115425} (\bibinfo {year}
  {2015})}\BibitemShut {NoStop}%
\bibitem [{\citenamefont {Long}\ and\ \citenamefont {Liu}(2016)}]{Long2016}%
  \BibitemOpen
  \bibfield  {author} {\bibinfo {author} {\bibfnamefont {R.}~\bibnamefont
  {Long}}\ and\ \bibinfo {author} {\bibfnamefont {W.}~\bibnamefont {Liu}},\
  }\bibfield  {title} {\bibinfo {title} {Efficiency and its bounds of minimally
  nonlinear irreversible heat engines at arbitrary power},\ }\href
  {https://doi.org/10.1103/PhysRevE.94.052114} {\bibfield  {journal} {\bibinfo
  {journal} {Phys. Rev. E}\ }\textbf {\bibinfo {volume} {94}},\ \bibinfo
  {pages} {052114} (\bibinfo {year} {2016})}\BibitemShut {NoStop}%
\bibitem [{\citenamefont {Holubec}\ and\ \citenamefont
  {Ryabov}(2016)}]{holubec2016maximum}%
  \BibitemOpen
  \bibfield  {author} {\bibinfo {author} {\bibfnamefont {V.}~\bibnamefont
  {Holubec}}\ and\ \bibinfo {author} {\bibfnamefont {A.}~\bibnamefont
  {Ryabov}},\ }\bibfield  {title} {\bibinfo {title} {Maximum efficiency of
  low-dissipation heat engines at arbitrary power},\ }\href@noop {} {\bibfield
  {journal} {\bibinfo  {journal} {J. Stat. Mech.}\ }\textbf {\bibinfo {volume}
  {2016}},\ \bibinfo {pages} {073204} (\bibinfo {year} {2016})}\BibitemShut
  {NoStop}%
\bibitem [{\citenamefont {Ye}\ and\ \citenamefont
  {Holubec}(2021)}]{ye2020maximum}%
  \BibitemOpen
  \bibfield  {author} {\bibinfo {author} {\bibfnamefont {Z.}~\bibnamefont
  {Ye}}\ and\ \bibinfo {author} {\bibfnamefont {V.}~\bibnamefont {Holubec}},\
  }\bibfield  {title} {\bibinfo {title} {Maximum efficiency of absorption
  refrigerators at arbitrary cooling power},\ }\href
  {https://doi.org/10.1103/PhysRevE.103.052125} {\bibfield  {journal} {\bibinfo
   {journal} {Phys. Rev. E}\ }\textbf {\bibinfo {volume} {103}},\ \bibinfo
  {pages} {052125} (\bibinfo {year} {2021})}\BibitemShut {NoStop}%
\bibitem [{\citenamefont {Guo}\ \emph {et~al.}(2020)\citenamefont {Guo},
  \citenamefont {Yang}, \citenamefont {Gonzalez-Ayala}, \citenamefont {Roco},
  \citenamefont {Medina},\ and\ \citenamefont
  {Hern{\'a}ndez}}]{guo2020equivalent}%
  \BibitemOpen
  \bibfield  {author} {\bibinfo {author} {\bibfnamefont {J.}~\bibnamefont
  {Guo}}, \bibinfo {author} {\bibfnamefont {H.}~\bibnamefont {Yang}}, \bibinfo
  {author} {\bibfnamefont {J.}~\bibnamefont {Gonzalez-Ayala}}, \bibinfo
  {author} {\bibfnamefont {J.}~\bibnamefont {Roco}}, \bibinfo {author}
  {\bibfnamefont {A.}~\bibnamefont {Medina}},\ and\ \bibinfo {author}
  {\bibfnamefont {A.~C.}\ \bibnamefont {Hern{\'a}ndez}},\ }\bibfield  {title}
  {\bibinfo {title} {The equivalent low-dissipation combined cycle system and
  optimal analyses of a class of thermally driven heat pumps},\ }\href@noop {}
  {\bibfield  {journal} {\bibinfo  {journal} {Energy Convers. Manag.}\ }\textbf
  {\bibinfo {volume} {220}},\ \bibinfo {pages} {113100} (\bibinfo {year}
  {2020})}\BibitemShut {NoStop}%
\bibitem [{\citenamefont {Cheng}\ and\ \citenamefont
  {Chen}(1995)}]{cheng1995performance}%
  \BibitemOpen
  \bibfield  {author} {\bibinfo {author} {\bibfnamefont {C.-Y.}\ \bibnamefont
  {Cheng}}\ and\ \bibinfo {author} {\bibfnamefont {C.-K.}\ \bibnamefont
  {Chen}},\ }\bibfield  {title} {\bibinfo {title} {Performance optimization of
  an irreversible heat pump},\ }\href@noop {} {\bibfield  {journal} {\bibinfo
  {journal} {J. Phys. D: Appl. Phys}\ }\textbf {\bibinfo {volume} {28}},\
  \bibinfo {pages} {2451} (\bibinfo {year} {1995})}\BibitemShut {NoStop}%
\bibitem [{\citenamefont {Chen}\ \emph {et~al.}(1995)\citenamefont {Chen},
  \citenamefont {Sun}, \citenamefont {Cheng},\ and\ \citenamefont
  {Chen}}]{chen1995study}%
  \BibitemOpen
  \bibfield  {author} {\bibinfo {author} {\bibfnamefont {W.}~\bibnamefont
  {Chen}}, \bibinfo {author} {\bibfnamefont {F.}~\bibnamefont {Sun}}, \bibinfo
  {author} {\bibfnamefont {S.}~\bibnamefont {Cheng}},\ and\ \bibinfo {author}
  {\bibfnamefont {L.}~\bibnamefont {Chen}},\ }\bibfield  {title} {\bibinfo
  {title} {Study on optimal performance and working temperatures of
  endoreversible forward and reverse carnot cycles},\ }\href@noop {} {\bibfield
   {journal} {\bibinfo  {journal} {Int. J. Energy Res.}\ }\textbf {\bibinfo
  {volume} {19}},\ \bibinfo {pages} {751} (\bibinfo {year} {1995})}\BibitemShut
  {NoStop}%
\bibitem [{\citenamefont {Salamon}\ \emph {et~al.}(1980)\citenamefont
  {Salamon}, \citenamefont {Nitzan}, \citenamefont {Andresen},\ and\
  \citenamefont {Berry}}]{PhysRevA.21.2115}%
  \BibitemOpen
  \bibfield  {author} {\bibinfo {author} {\bibfnamefont {P.}~\bibnamefont
  {Salamon}}, \bibinfo {author} {\bibfnamefont {A.}~\bibnamefont {Nitzan}},
  \bibinfo {author} {\bibfnamefont {B.}~\bibnamefont {Andresen}},\ and\
  \bibinfo {author} {\bibfnamefont {R.~S.}\ \bibnamefont {Berry}},\ }\bibfield
  {title} {\bibinfo {title} {Minimum entropy production and the optimization of
  heat engines},\ }\href {https://doi.org/10.1103/PhysRevA.21.2115} {\bibfield
  {journal} {\bibinfo  {journal} {Phys. Rev. A}\ }\textbf {\bibinfo {volume}
  {21}},\ \bibinfo {pages} {2115} (\bibinfo {year} {1980})}\BibitemShut
  {NoStop}%
\bibitem [{\citenamefont {Gerstenmaier}(2021)}]{PhysRevE.103.032141}%
  \BibitemOpen
  \bibfield  {author} {\bibinfo {author} {\bibfnamefont {Y.~C.}\ \bibnamefont
  {Gerstenmaier}},\ }\bibfield  {title} {\bibinfo {title} {Irreversible entropy
  production in low- and high-dissipation heat engines and the problem of the
  curzon-ahlborn efficiency},\ }\href
  {https://doi.org/10.1103/PhysRevE.103.032141} {\bibfield  {journal} {\bibinfo
   {journal} {Phys. Rev. E}\ }\textbf {\bibinfo {volume} {103}},\ \bibinfo
  {pages} {032141} (\bibinfo {year} {2021})}\BibitemShut {NoStop}%
\bibitem [{\citenamefont {Sekimoto}\ and\ \citenamefont
  {Sasa}(1997)}]{Sekimoto1997}%
  \BibitemOpen
  \bibfield  {author} {\bibinfo {author} {\bibfnamefont {K.}~\bibnamefont
  {Sekimoto}}\ and\ \bibinfo {author} {\bibfnamefont {S.-i.}\ \bibnamefont
  {Sasa}},\ }\bibfield  {title} {\bibinfo {title} {Complementarity relation for
  irreversible process derived from stochastic energetics},\ }\href
  {https://doi.org/10.1143/JPSJ.66.3326} {\bibfield  {journal} {\bibinfo
  {journal} {J. Phys. Soc. Japan}\ }\textbf {\bibinfo {volume} {66}},\ \bibinfo
  {pages} {3326} (\bibinfo {year} {1997})}\BibitemShut {NoStop}%
\bibitem [{\citenamefont {Zulkowski}\ and\ \citenamefont
  {DeWeese}(2015)}]{Zulkowski2015}%
  \BibitemOpen
  \bibfield  {author} {\bibinfo {author} {\bibfnamefont {P.~R.}\ \bibnamefont
  {Zulkowski}}\ and\ \bibinfo {author} {\bibfnamefont {M.~R.}\ \bibnamefont
  {DeWeese}},\ }\bibfield  {title} {\bibinfo {title} {Optimal protocols for
  slowly driven quantum systems},\ }\href
  {https://doi.org/10.1103/PhysRevE.92.032113} {\bibfield  {journal} {\bibinfo
  {journal} {Phys. Rev. E}\ }\textbf {\bibinfo {volume} {92}},\ \bibinfo
  {pages} {032113} (\bibinfo {year} {2015})}\BibitemShut {NoStop}%
\bibitem [{\citenamefont {Cavina}\ \emph {et~al.}(2017)\citenamefont {Cavina},
  \citenamefont {Mari},\ and\ \citenamefont {Giovannetti}}]{Cavina2017}%
  \BibitemOpen
  \bibfield  {author} {\bibinfo {author} {\bibfnamefont {V.}~\bibnamefont
  {Cavina}}, \bibinfo {author} {\bibfnamefont {A.}~\bibnamefont {Mari}},\ and\
  \bibinfo {author} {\bibfnamefont {V.}~\bibnamefont {Giovannetti}},\
  }\bibfield  {title} {\bibinfo {title} {Slow dynamics and thermodynamics of
  open quantum systems},\ }\href
  {https://doi.org/10.1103/PhysRevLett.119.050601} {\bibfield  {journal}
  {\bibinfo  {journal} {Phys. Rev. Lett.}\ }\textbf {\bibinfo {volume} {119}},\
  \bibinfo {pages} {050601} (\bibinfo {year} {2017})}\BibitemShut {NoStop}%
\bibitem [{\citenamefont {Ma}\ \emph {et~al.}(2020)\citenamefont {Ma},
  \citenamefont {Zhai}, \citenamefont {Chen}, \citenamefont {Sun},\ and\
  \citenamefont {Dong}}]{Ma/etal:2020}%
  \BibitemOpen
  \bibfield  {author} {\bibinfo {author} {\bibfnamefont {Y.-H.}\ \bibnamefont
  {Ma}}, \bibinfo {author} {\bibfnamefont {R.-X.}\ \bibnamefont {Zhai}},
  \bibinfo {author} {\bibfnamefont {J.}~\bibnamefont {Chen}}, \bibinfo {author}
  {\bibfnamefont {C.~P.}\ \bibnamefont {Sun}},\ and\ \bibinfo {author}
  {\bibfnamefont {H.}~\bibnamefont {Dong}},\ }\bibfield  {title} {\bibinfo
  {title} {Experimental test of the $1/\ensuremath{\tau}$-scaling entropy
  generation in finite-time thermodynamics},\ }\href
  {https://doi.org/10.1103/PhysRevLett.125.210601} {\bibfield  {journal}
  {\bibinfo  {journal} {Phys. Rev. Lett.}\ }\textbf {\bibinfo {volume} {125}},\
  \bibinfo {pages} {210601} (\bibinfo {year} {2020})}\BibitemShut {NoStop}%
\bibitem [{\citenamefont {Holubec}\ \emph {et~al.}(2020)\citenamefont
  {Holubec}, \citenamefont {Steffenoni}, \citenamefont {Falasco},\ and\
  \citenamefont {Kroy}}]{Holubec2020}%
  \BibitemOpen
  \bibfield  {author} {\bibinfo {author} {\bibfnamefont {V.}~\bibnamefont
  {Holubec}}, \bibinfo {author} {\bibfnamefont {S.}~\bibnamefont {Steffenoni}},
  \bibinfo {author} {\bibfnamefont {G.}~\bibnamefont {Falasco}},\ and\ \bibinfo
  {author} {\bibfnamefont {K.}~\bibnamefont {Kroy}},\ }\bibfield  {title}
  {\bibinfo {title} {Active brownian heat engines},\ }\href
  {https://doi.org/10.1103/PhysRevResearch.2.043262} {\bibfield  {journal}
  {\bibinfo  {journal} {Phys. Rev. Res.}\ }\textbf {\bibinfo {volume} {2}},\
  \bibinfo {pages} {043262} (\bibinfo {year} {2020})}\BibitemShut {NoStop}%
\bibitem [{\citenamefont {Iyyappan}\ and\ \citenamefont
  {Johal}(2020)}]{iyyappan2020efficiency}%
  \BibitemOpen
  \bibfield  {author} {\bibinfo {author} {\bibfnamefont {I.}~\bibnamefont
  {Iyyappan}}\ and\ \bibinfo {author} {\bibfnamefont {R.~S.}\ \bibnamefont
  {Johal}},\ }\bibfield  {title} {\bibinfo {title} {Efficiency of a two-stage
  heat engine at optimal power},\ }\href@noop {} {\bibfield  {journal}
  {\bibinfo  {journal} {EPL}\ }\textbf {\bibinfo {volume} {128}},\ \bibinfo
  {pages} {50004} (\bibinfo {year} {2020})}\BibitemShut {NoStop}%
\bibitem [{\citenamefont {Blickle}\ and\ \citenamefont
  {Bechinger}(2012)}]{blickle2012realization}%
  \BibitemOpen
  \bibfield  {author} {\bibinfo {author} {\bibfnamefont {V.}~\bibnamefont
  {Blickle}}\ and\ \bibinfo {author} {\bibfnamefont {C.}~\bibnamefont
  {Bechinger}},\ }\bibfield  {title} {\bibinfo {title} {Realization of a
  micrometre-sized stochastic heat engine},\ }\href@noop {} {\bibfield
  {journal} {\bibinfo  {journal} {Nat. Phys.}\ }\textbf {\bibinfo {volume}
  {8}},\ \bibinfo {pages} {143} (\bibinfo {year} {2012})}\BibitemShut {NoStop}%
\bibitem [{\citenamefont {Ahmadi}\ \emph {et~al.}(2015)\citenamefont {Ahmadi},
  \citenamefont {Ahmadi}, \citenamefont {Mehrpooya},\ and\ \citenamefont
  {Sameti}}]{ahmadi2015thermo}%
  \BibitemOpen
  \bibfield  {author} {\bibinfo {author} {\bibfnamefont {M.~H.}\ \bibnamefont
  {Ahmadi}}, \bibinfo {author} {\bibfnamefont {M.~A.}\ \bibnamefont {Ahmadi}},
  \bibinfo {author} {\bibfnamefont {M.}~\bibnamefont {Mehrpooya}},\ and\
  \bibinfo {author} {\bibfnamefont {M.}~\bibnamefont {Sameti}},\ }\bibfield
  {title} {\bibinfo {title} {Thermo-ecological analysis and optimization
  performance of an irreversible three-heat-source absorption heat pump},\
  }\href@noop {} {\bibfield  {journal} {\bibinfo  {journal} {Energy Convers.
  Manag.}\ }\textbf {\bibinfo {volume} {90}},\ \bibinfo {pages} {175} (\bibinfo
  {year} {2015})}\BibitemShut {NoStop}%
\bibitem [{\citenamefont {Novikov}(1958)}]{Novikov1958}%
  \BibitemOpen
  \bibfield  {author} {\bibinfo {author} {\bibfnamefont {I.~I.}\ \bibnamefont
  {Novikov}},\ }\bibfield  {title} {\bibinfo {title} {{The efficiency of atomic
  power stations}},\ }\href@noop {} {\bibfield  {journal} {\bibinfo  {journal}
  {J. Nucl. Energy II}\ }\textbf {\bibinfo {volume} {7}},\ \bibinfo {pages}
  {125} (\bibinfo {year} {1958})}\BibitemShut {NoStop}%
\bibitem [{\citenamefont {Chambadal}(1957)}]{Chambadal1957}%
  \BibitemOpen
  \bibfield  {author} {\bibinfo {author} {\bibfnamefont {P.}~\bibnamefont
  {Chambadal}},\ }\href@noop {} {\emph {\bibinfo {title} {Les centrales
  nucl{\'e}aires}}},\ Vol.\ \bibinfo {volume} {321}\ (\bibinfo  {publisher}
  {Colin},\ \bibinfo {year} {1957})\BibitemShut {NoStop}%
\bibitem [{\citenamefont {Curzon}\ and\ \citenamefont
  {Ahlborn}(1975)}]{curzon1975efficiency}%
  \BibitemOpen
  \bibfield  {author} {\bibinfo {author} {\bibfnamefont {F.~L.}\ \bibnamefont
  {Curzon}}\ and\ \bibinfo {author} {\bibfnamefont {B.}~\bibnamefont
  {Ahlborn}},\ }\bibfield  {title} {\bibinfo {title} {Efficiency of a carnot
  engine at maximum power output},\ }\href@noop {} {\bibfield  {journal}
  {\bibinfo  {journal} {Am. J. Phys.}\ }\textbf {\bibinfo {volume} {43}},\
  \bibinfo {pages} {22} (\bibinfo {year} {1975})}\BibitemShut {NoStop}%
\bibitem [{\citenamefont {Chen}\ \emph {et~al.}(2021)\citenamefont {Chen},
  \citenamefont {Chen}, \citenamefont {Fei},\ and\ \citenamefont
  {Quan}}]{chen2021microscopic}%
  \BibitemOpen
  \bibfield  {author} {\bibinfo {author} {\bibfnamefont {Y.}~\bibnamefont
  {Chen}}, \bibinfo {author} {\bibfnamefont {J.-F.}\ \bibnamefont {Chen}},
  \bibinfo {author} {\bibfnamefont {Z.}~\bibnamefont {Fei}},\ and\ \bibinfo
  {author} {\bibfnamefont {H.}~\bibnamefont {Quan}},\ }\bibfield  {title}
  {\bibinfo {title} {A microscopic theory of curzon-ahlborn heat engine},\
  }\href@noop {} {\bibfield  {journal} {\bibinfo  {journal} {arXiv preprint
  arXiv:2108.04128}\ } (\bibinfo {year} {2021})}\BibitemShut {NoStop}%
\bibitem [{\citenamefont {Vaudrey}\ \emph {et~al.}(2009)\citenamefont
  {Vaudrey}, \citenamefont {Baucour}, \citenamefont {Lanzetta},\ and\
  \citenamefont {Glises}}]{vaudrey2009detailed}%
  \BibitemOpen
  \bibfield  {author} {\bibinfo {author} {\bibfnamefont {A.}~\bibnamefont
  {Vaudrey}}, \bibinfo {author} {\bibfnamefont {P.}~\bibnamefont {Baucour}},
  \bibinfo {author} {\bibfnamefont {F.}~\bibnamefont {Lanzetta}},\ and\
  \bibinfo {author} {\bibfnamefont {R.}~\bibnamefont {Glises}},\ }\bibfield
  {title} {\bibinfo {title} {Detailed analysis of an endoreversible fuel cell:
  Maximum power and optimal operating temperature determination},\ }\href@noop
  {} {\bibfield  {journal} {\bibinfo  {journal} {arXiv preprint
  arXiv:0905.2871}\ } (\bibinfo {year} {2009})}\BibitemShut {NoStop}%
\bibitem [{\citenamefont {Bouton}\ \emph {et~al.}(2021)\citenamefont {Bouton},
  \citenamefont {Nettersheim}, \citenamefont {Burgardt}, \citenamefont {Adam},
  \citenamefont {Lutz},\ and\ \citenamefont {Widera}}]{bouton2021quantum}%
  \BibitemOpen
  \bibfield  {author} {\bibinfo {author} {\bibfnamefont {Q.}~\bibnamefont
  {Bouton}}, \bibinfo {author} {\bibfnamefont {J.}~\bibnamefont {Nettersheim}},
  \bibinfo {author} {\bibfnamefont {S.}~\bibnamefont {Burgardt}}, \bibinfo
  {author} {\bibfnamefont {D.}~\bibnamefont {Adam}}, \bibinfo {author}
  {\bibfnamefont {E.}~\bibnamefont {Lutz}},\ and\ \bibinfo {author}
  {\bibfnamefont {A.}~\bibnamefont {Widera}},\ }\bibfield  {title} {\bibinfo
  {title} {A quantum heat engine driven by atomic collisions},\ }\href@noop {}
  {\bibfield  {journal} {\bibinfo  {journal} {Nat. Commun.}\ }\textbf {\bibinfo
  {volume} {12}},\ \bibinfo {pages} {1} (\bibinfo {year} {2021})}\BibitemShut
  {NoStop}%
\bibitem [{\citenamefont {Chen}\ and\ \citenamefont
  {Yan}(1989{\natexlab{b}})}]{chen1989effect}%
  \BibitemOpen
  \bibfield  {author} {\bibinfo {author} {\bibfnamefont {L.}~\bibnamefont
  {Chen}}\ and\ \bibinfo {author} {\bibfnamefont {Z.}~\bibnamefont {Yan}},\
  }\bibfield  {title} {\bibinfo {title} {The effect of heat-transfer law on
  performance of a two-heat-source endoreversible cycle},\ }\href@noop {}
  {\bibfield  {journal} {\bibinfo  {journal} {J. Chem. Phys.}\ }\textbf
  {\bibinfo {volume} {90}},\ \bibinfo {pages} {3740} (\bibinfo {year}
  {1989}{\natexlab{b}})}\BibitemShut {NoStop}%
\bibitem [{\citenamefont {Huleihil}\ and\ \citenamefont
  {Andresen}(2006)}]{huleihil2006convective}%
  \BibitemOpen
  \bibfield  {author} {\bibinfo {author} {\bibfnamefont {M.}~\bibnamefont
  {Huleihil}}\ and\ \bibinfo {author} {\bibfnamefont {B.}~\bibnamefont
  {Andresen}},\ }\bibfield  {title} {\bibinfo {title} {Convective heat transfer
  law for an endoreversible engine},\ }\href@noop {} {\bibfield  {journal}
  {\bibinfo  {journal} {J. Appl. Phys.}\ }\textbf {\bibinfo {volume} {100}},\
  \bibinfo {pages} {014911} (\bibinfo {year} {2006})}\BibitemShut {NoStop}%
\bibitem [{\citenamefont {Abiuso}\ and\ \citenamefont
  {Perarnau-Llobet}(2020)}]{PhysRevLett.124.110606}%
  \BibitemOpen
  \bibfield  {author} {\bibinfo {author} {\bibfnamefont {P.}~\bibnamefont
  {Abiuso}}\ and\ \bibinfo {author} {\bibfnamefont {M.}~\bibnamefont
  {Perarnau-Llobet}},\ }\bibfield  {title} {\bibinfo {title} {Optimal cycles
  for low-dissipation heat engines},\ }\href
  {https://doi.org/10.1103/PhysRevLett.124.110606} {\bibfield  {journal}
  {\bibinfo  {journal} {Phys. Rev. Lett.}\ }\textbf {\bibinfo {volume} {124}},\
  \bibinfo {pages} {110606} (\bibinfo {year} {2020})}\BibitemShut {NoStop}%
\bibitem [{\citenamefont {Abiuso}\ \emph {et~al.}(2020)\citenamefont {Abiuso},
  \citenamefont {JD~Miller}, \citenamefont {Perarnau-Llobet},\ and\
  \citenamefont {Scandi}}]{abiuso2020geometric}%
  \BibitemOpen
  \bibfield  {author} {\bibinfo {author} {\bibfnamefont {P.}~\bibnamefont
  {Abiuso}}, \bibinfo {author} {\bibfnamefont {H.}~\bibnamefont {JD~Miller}},
  \bibinfo {author} {\bibfnamefont {M.}~\bibnamefont {Perarnau-Llobet}},\ and\
  \bibinfo {author} {\bibfnamefont {M.}~\bibnamefont {Scandi}},\ }\bibfield
  {title} {\bibinfo {title} {Geometric optimisation of quantum thermodynamic
  processes},\ }\href@noop {} {\bibfield  {journal} {\bibinfo  {journal}
  {Entropy}\ }\textbf {\bibinfo {volume} {22}},\ \bibinfo {pages} {1076}
  (\bibinfo {year} {2020})}\BibitemShut {NoStop}%
\bibitem [{\citenamefont {Reyes-Ram\'{\i}rez}\ \emph
  {et~al.}(2017)\citenamefont {Reyes-Ram\'{\i}rez}, \citenamefont
  {Gonzalez-Ayala}, \citenamefont {Calvo~Hern\'andez},\ and\ \citenamefont
  {Santill\'an}}]{PhysRevE.96.042128}%
  \BibitemOpen
  \bibfield  {author} {\bibinfo {author} {\bibfnamefont {I.}~\bibnamefont
  {Reyes-Ram\'{\i}rez}}, \bibinfo {author} {\bibfnamefont {J.}~\bibnamefont
  {Gonzalez-Ayala}}, \bibinfo {author} {\bibfnamefont {A.}~\bibnamefont
  {Calvo~Hern\'andez}},\ and\ \bibinfo {author} {\bibfnamefont
  {M.}~\bibnamefont {Santill\'an}},\ }\bibfield  {title} {\bibinfo {title}
  {Local-stability analysis of a low-dissipation heat engine working at maximum
  power output},\ }\href {https://doi.org/10.1103/PhysRevE.96.042128}
  {\bibfield  {journal} {\bibinfo  {journal} {Phys. Rev. E}\ }\textbf {\bibinfo
  {volume} {96}},\ \bibinfo {pages} {042128} (\bibinfo {year}
  {2017})}\BibitemShut {NoStop}%
\bibitem [{\citenamefont {Gonzalez-Ayala}\ \emph {et~al.}(2018)\citenamefont
  {Gonzalez-Ayala}, \citenamefont {Santill\'an}, \citenamefont
  {Reyes-Ram\'{\i}rez},\ and\ \citenamefont
  {Calvo-Hern\'andez}}]{PhysRevE.98.032142}%
  \BibitemOpen
  \bibfield  {author} {\bibinfo {author} {\bibfnamefont {J.}~\bibnamefont
  {Gonzalez-Ayala}}, \bibinfo {author} {\bibfnamefont {M.}~\bibnamefont
  {Santill\'an}}, \bibinfo {author} {\bibfnamefont {I.}~\bibnamefont
  {Reyes-Ram\'{\i}rez}},\ and\ \bibinfo {author} {\bibfnamefont
  {A.}~\bibnamefont {Calvo-Hern\'andez}},\ }\bibfield  {title} {\bibinfo
  {title} {Link between optimization and local stability of a low-dissipation
  heat engine: Dynamic and energetic behaviors},\ }\href
  {https://doi.org/10.1103/PhysRevE.98.032142} {\bibfield  {journal} {\bibinfo
  {journal} {Phys. Rev. E}\ }\textbf {\bibinfo {volume} {98}},\ \bibinfo
  {pages} {032142} (\bibinfo {year} {2018})}\BibitemShut {NoStop}%
\bibitem [{\citenamefont {Gonzalez-Ayala}\ \emph {et~al.}(2017)\citenamefont
  {Gonzalez-Ayala}, \citenamefont {Calvo~Hern\'andez},\ and\ \citenamefont
  {Roco}}]{PhysRevE.95.022131}%
  \BibitemOpen
  \bibfield  {author} {\bibinfo {author} {\bibfnamefont {J.}~\bibnamefont
  {Gonzalez-Ayala}}, \bibinfo {author} {\bibfnamefont {A.}~\bibnamefont
  {Calvo~Hern\'andez}},\ and\ \bibinfo {author} {\bibfnamefont {J.~M.~M.}\
  \bibnamefont {Roco}},\ }\bibfield  {title} {\bibinfo {title} {From maximum
  power to a trade-off optimization of low-dissipation heat engines: Influence
  of control parameters and the role of entropy generation},\ }\href
  {https://doi.org/10.1103/PhysRevE.95.022131} {\bibfield  {journal} {\bibinfo
  {journal} {Phys. Rev. E}\ }\textbf {\bibinfo {volume} {95}},\ \bibinfo
  {pages} {022131} (\bibinfo {year} {2017})}\BibitemShut {NoStop}%
\bibitem [{\citenamefont {Ondrechen}\ \emph {et~al.}(1983)\citenamefont
  {Ondrechen}, \citenamefont {Rubin},\ and\ \citenamefont
  {Band}}]{ondrechen1983generalized}%
  \BibitemOpen
  \bibfield  {author} {\bibinfo {author} {\bibfnamefont {M.~J.}\ \bibnamefont
  {Ondrechen}}, \bibinfo {author} {\bibfnamefont {M.~H.}\ \bibnamefont
  {Rubin}},\ and\ \bibinfo {author} {\bibfnamefont {Y.~B.}\ \bibnamefont
  {Band}},\ }\bibfield  {title} {\bibinfo {title} {The generalized carnot
  cycle: A working fluid operating in finite time between finite heat sources
  and sinks},\ }\href@noop {} {\bibfield  {journal} {\bibinfo  {journal} {J.
  Chem. Phys.}\ }\textbf {\bibinfo {volume} {78}},\ \bibinfo {pages} {4721}
  (\bibinfo {year} {1983})}\BibitemShut {NoStop}%
\bibitem [{\citenamefont {Wang}(2014)}]{PhysRevE.90.062140}%
  \BibitemOpen
  \bibfield  {author} {\bibinfo {author} {\bibfnamefont {Y.}~\bibnamefont
  {Wang}},\ }\bibfield  {title} {\bibinfo {title} {Optimization in
  finite-reservoir finite-time thermodynamics},\ }\href
  {https://doi.org/10.1103/PhysRevE.90.062140} {\bibfield  {journal} {\bibinfo
  {journal} {Phys. Rev. E}\ }\textbf {\bibinfo {volume} {90}},\ \bibinfo
  {pages} {062140} (\bibinfo {year} {2014})}\BibitemShut {NoStop}%
\bibitem [{\citenamefont {Wang}(2016)}]{PhysRevE.93.012120}%
  \BibitemOpen
  \bibfield  {author} {\bibinfo {author} {\bibfnamefont {Y.}~\bibnamefont
  {Wang}},\ }\bibfield  {title} {\bibinfo {title} {Optimizing work output for
  finite-sized heat reservoirs: Beyond linear response},\ }\href
  {https://doi.org/10.1103/PhysRevE.93.012120} {\bibfield  {journal} {\bibinfo
  {journal} {Phys. Rev. E}\ }\textbf {\bibinfo {volume} {93}},\ \bibinfo
  {pages} {012120} (\bibinfo {year} {2016})}\BibitemShut {NoStop}%
\bibitem [{\citenamefont {Ma}(2020)}]{ma2020effect}%
  \BibitemOpen
  \bibfield  {author} {\bibinfo {author} {\bibfnamefont {Y.-H.}\ \bibnamefont
  {Ma}},\ }\bibfield  {title} {\bibinfo {title} {Effect of finite-size heat
  source’s heat capacity on the efficiency of heat engine},\ }\href@noop {}
  {\bibfield  {journal} {\bibinfo  {journal} {Entropy}\ }\textbf {\bibinfo
  {volume} {22}},\ \bibinfo {pages} {1002} (\bibinfo {year}
  {2020})}\BibitemShut {NoStop}%
\bibitem [{\citenamefont {Yuan}\ \emph {et~al.}(2021)\citenamefont {Yuan},
  \citenamefont {Ma},\ and\ \citenamefont {Sun}}]{yuan2021optimizing}%
  \BibitemOpen
  \bibfield  {author} {\bibinfo {author} {\bibfnamefont {H.}~\bibnamefont
  {Yuan}}, \bibinfo {author} {\bibfnamefont {Y.-H.}\ \bibnamefont {Ma}},\ and\
  \bibinfo {author} {\bibfnamefont {C.}~\bibnamefont {Sun}},\ }\bibfield
  {title} {\bibinfo {title} {Optimizing thermodynamic cycles with two
  finite-sized reservoirs},\ }\href@noop {} {\bibfield  {journal} {\bibinfo
  {journal} {arXiv preprint arXiv:2107.11342}\ } (\bibinfo {year}
  {2021})}\BibitemShut {NoStop}%
\end{thebibliography}%

\end{document}